\documentclass[nofootinbib,superscriptaddress,a4paper, twocolumn, floatfix]{revtex4-2}
\usepackage[english]{babel}
\usepackage[utf8]{inputenc}
\usepackage[colorinlistoftodos, color=green!40, prependcaption]{todonotes}
\usepackage{amsthm}
\usepackage{mathtools}
\usepackage{physics}
\usepackage{braket}
\usepackage{listings}
\usepackage{xcolor}
\usepackage{graphicx}
\usepackage[left=23mm,right=13mm,top=35mm,columnsep=15pt]{geometry} 
\usepackage{adjustbox}
\usepackage{soul}
\usepackage{placeins}
\usepackage[T1]{fontenc}
\usepackage{lipsum}
\usepackage{csquotes}
\usepackage[pdftex, pdftitle={Article}, pdfauthor={Author}]{hyperref} 
\usepackage{booktabs}
\usepackage{amssymb}
\usepackage{bm}
\usepackage[ruled,vlined]{algorithm2e}
\usepackage[caption=false]{subfig}
\usepackage{diagbox}

\theoremstyle{plain} 

\theoremstyle{definition}

\theoremstyle{remark}

\newcommand{\mycomment}[1]{}
\begin{document}
\title{Dynamic local single-shot checks for toric codes}

\author{Yingjia Lin}
\email[E-mail:]{yingjia.lin@duke.edu}
\affiliation{Duke Quantum Center, Duke University, Durham, NC 27701, USA.}
\affiliation{Department of Physics, Duke University, Durham, NC 27708, USA.}

\author{Abhinav Anand}
\email[E-mail:]{abhinav.anand@duke.edu}
\affiliation{Duke Quantum Center, Duke University, Durham, NC 27701, USA.}
\affiliation{Department of Electrical and Computer Engineering, Duke University, Durham, NC 27708, USA.}

\author{Kenneth R. Brown}
\email[E-mail:]{kenneth.r.brown@duke.edu}
\affiliation{Duke Quantum Center, Duke University, Durham, NC 27701, USA.}
\affiliation{Department of Physics, Duke University, Durham, NC 27708, USA.}
\affiliation{Department of Electrical and Computer Engineering, Duke University, Durham, NC 27708, USA.}
\affiliation{Department of Chemistry, Duke University, Durham, NC 27708, USA.}
\date{\today}

\begin{abstract}
Quantum error correction typically requires repeated syndrome extraction due to measurement noise, which results in substantial time overhead in fault-tolerant computation. 
Single-shot error correction aims to suppress errors using only one round of syndrome extraction. 
However, for most codes, it requires high-weight checks, which significantly degrade, and often eliminate, single-shot performance at the circuit level. 
In this work, we introduce local single-shot checks, where we impose constraints on check weights. 
Using a dynamic measurement scheme, we show that the number of required measurement rounds can be reduced by a factor determined by this constraint. 
As an example, we show through numerical simulation that our scheme can improve decoding performance compared to conventional checks when using sliding-window decoding with a reduced window size under circuit-level noise models for toric codes. 
Our work provides a new direction for constructing checks that can reduce time overhead in large-scale fault-tolerant quantum computation.
\end{abstract}

\maketitle

\section{Introduction}

Quantum error correction is a promising approach to achieve the low logical error rates required for large-scale quantum computation~\cite{calderbank1996good,PhysRevLett.77.793,gottesman2009introductionquantumerrorcorrection,RevModPhys.87.307,campbell2024series}. Recent experiments~\cite{google2025quantum,PRXQuantum.5.030326,PhysRevX.11.041058,bluvstein2024logical, EganNature2021} have demonstrated significant progress towards realizing fault-tolerant quantum computation. 
Quantum error correction uses syndrome measurements to detect and correct errors.
However, since the measurement process itself is prone to errors \cite{google2025quantum,PhysRevLett.134.100601,PhysRevX.13.041052,wu2019stern,PhysRevLett.129.203602}, multiple rounds of measurement, scaled with the code distance, are typically required within each decoding window \cite{dennis2002topological,huang2021constructions,huang2021between,tan2023scalable,skoric2023parallel,PhysRevA.86.032324}.
This leads to a significant time overhead in fault-tolerant quantum computation \cite{PRXQuantum.6.020326,mcardle2025fastcuriousacceleratefaulttolerant,PhysRevA.86.032324,horsman2012surface}.

Reducing the number of measurement rounds per decoding window has therefore been a focus of recent work. 
In particular, single-shot error correction aims to suppress errors using only one round of syndrome extraction \cite{PhysRevX.5.031043,campbell2019theory,PRXQuantum.2.020340,brown2016fault,kubica2022single}. 
Although single-shot checks can be defined for general codes \cite{campbell2019theory}, they are often high-weight, which leads to degraded circuit-level performance \cite{lin2024single}. Data-syndrome codes \cite{ashikhmin2014robust,ashikhmin2016correction,ashikhmin2020quantum,PhysRevA.90.062304} provide an alternative approach by encoding the measured syndrome using a classical code, with the required number of measurement repetitions depending on the classical code length. However, these constructions also typically involve high-weight checks, which are expected to similarly degrade circuit-level performance \cite{anker2025compressingsyndromemeasurementsequences}.

Another approach is to use encoded syndrome qubits to reduce measurement repetition. 
In Steane \cite{steane1997} or Knill error correction \cite{knill2005quantum}, a single round of measurement is sufficient, at the cost of encoding the syndrome qubits using the same code as the data qubits. 
Variants using syndrome qubits encoded with intermediate-size codes reduce the measurement overhead by a constant factor proportional to the encoding size, as shown for toric codes in Refs.~\cite{huang2021between,huang2021constructions}. 
Encoding syndrome qubits can also eliminate certain hook errors \cite{96shor,divincenzo1996fault}. 
However, preparing high-fidelity encoded syndrome states remains challenging, even for relatively simple states such as cat states \cite{PhysRevLett.98.020501,prabhu2023fault,knill2005quantum,reichardt2004improved,steane2004fastfaulttolerantfilteringquantum}.

In particular, the construction in Refs.~\cite{huang2021between,huang2021constructions} shows that measurement reduction can be achieved when the time distance within the decoding window is maintained at the code distance.
The authors maintain the time distance using offset alignment of encoded syndrome qubits across measurement rounds. 

The concept of time distance also inspires the construction introduced in this work.
Here, we propose local single-shot checks. 
This construction limits check weight by partitioning the code into overlapping qubit patches and constructing checks whose support lies within a single patch.
Using a dynamic measurement scheme, we increase the time distance achievable within a fixed number of syndrome extraction rounds, thereby reducing the number of measurement rounds required per decoding window. 
Previous dynamic measurement schemes have been used to reduce check weight or increase the number of logical qubits.~\cite{hastings2021dynamically,haah2022boundaries,PRXQuantum.4.020341} 
However, most schemes modify the logical code space during the measurement process, whereas our approach alternates between check sets that define the same code.
The effectiveness of our dynamic scheme arises from the tailored space–time structure of the decoding graph.

Numerical results show that our dynamic measurement scheme using local single-shot checks reduces measurement overhead by a constant factor determined by the patch size, while yielding an increased threshold under phenomenological noise. 
Although circuit-level behavior is affected by increased check weight and other circuit-level error sources \cite{delfosse2021beyond,fowler2010surface,PhysRevA.90.062320}, we provide circuit-level simulations demonstrating a threshold.
However, the performance of our checks is suboptimal compared to that of standard toric code checks in most cases considered here, but in certain parameter regimes, particularly when using sliding-window decoding with a reduced window size, our checks achieve performance better than standard toric code checks. 
We believe our approach offers a new pathway toward reducing time overhead in quantum error correction, with potential generalizations to other code families.

This paper is structured as follows: In Sec.~\ref{sec: background}, we review the basics of toric codes and different methods to reduce syndrome extraction rounds required for each decoding window. 
In Sec.~\ref{sec: checks}, we introduce the local single-shot we construct and the dynamic syndrome extraction scheme that achieves the measurement reduction. 
In Sec.~\ref{sec:circuit}, we introduce a syndrome extraction circuit to implement the local single-shot checks. 
We describe the noise models we used for simulation in Sec.~\ref{sec:noise_model} and construct decoders for each noise model in Sec.~\ref{sec: decoder}. 
We present our simulation results in Sec.~\ref{sec: results}. 
Finally, we conclude the paper and discuss potential future directions in Sec.~\ref{sec: conclusion}.

\section{Background}\label{sec: background}

\subsection{Toric code}
Stabilizer codes encode logical information in the $+1$ eigenspace of an abelian subgroup $S$ of the Pauli group. 
This abelian group $S$ is called the stabilizer group of the code. Each of its elements is a stabilizer.
Since errors can flip the outcomes of stabilizer measurements, measuring a generating set of stabilizers allows errors to be detected and corrected.
The operators selected for measurement are called checks.
The choice of checks is not unique, just as generators are not unique. 

The codes considered in this work are Calderbank–Shor–Steane (CSS) codes \cite{calderbank1996good,steane1996multiple}. 
For any CSS code, the checks can be chosen to be either Pauli-$X$ or Pauli-$Z$ operators, forming $X$-type and $Z$-type check sets.
The $Z$($X$)-type checks can be represented by $Z$($X$)-type parity check matrix $H_Z$($H_X$), where each element $H_Z^{ij}=1$ ($H_X^{ij}=1$) if the $i$th $Z$($X$)-type check has support on $j$th data qubit.
The abelian property of the check requires $H_Z H_X^T = 0$. 
Because $X$ errors only affect $Z$-type checks and vice versa, $X$ and $Z$ errors can be decoded independently.

We focus on the toric code in this paper. The toric code is a CSS code defined on an $L \times L$ square lattice with periodic boundary conditions \cite{kitaev2003fault}. 
It encodes two logical qubits in $2L^2$ physical qubits, where qubits lie on the edges of the lattice.
The $X$-type checks are defined at each vertex with support from qubits around the vertex. 
The $Z$-type checks are defined at each plaquette with support from qubits surrounding the plaquette.
We refer to this set of checks as the local checks, as in Ref. \cite{lin2024single}.

The code distance $d$ sets an upper bound on the weight of errors that can be reliably corrected in the absence of measurement errors. 
Specifically, errors of weight up to $ \lfloor (d-1)/2 \rfloor $
are all correctable. 
Here, the weight of an error denotes the number of physical qubits on which it has nontrivial support. 
For the toric code, the distance is $d=L$ and logical operators correspond to non-contractible loops wrapping around the torus.

To decode the toric code under repeated measurement, a decoding graph is constructed from detectors $D_{i,j}(t)$, where: $D_{i,j}(t) = M_{i,j}(t)M_{i,j}(t+1)$ is defined as the product of measurements of the same check from two consecutive syndrome extraction rounds, with the first round of detector defined as $D_{i,j}(0)=M_{i,j}(0)$.
Here $M_{i,j}(t)$ denotes the syndrome of the check at location $(i,j)$ on the torus at syndrome extraction round $t$. 
The label $(i,j)$ of the detector $D_{i,j}(t)$ therefore can be interpreted as the spatial position of the detector in the decoding graph. 
By a round of syndrome extraction, we mean extracting the syndrome from all the checks at the time step. 
Each detector is represented as a vertex in the decoding graph.
Errors correspond to edges (or hyperedges) connecting the detectors they flip. 
Since each local error flips at most two detectors, the toric code can be decoded using minimum-weight perfect matching (MWPM) \cite{edmonds_1965,edmonds1965maximum,fowler2010surface}.

\subsection{Measurement repetition under measurement errors}
In the presence of measurement noise, a single round of noisy syndrome data is typically insufficient for reliable decoding, and repeated measurements are required \cite{dennis2002topological,huang2021constructions,huang2021between}. For large-scale fault-tolerant computation, a common strategy is sliding-window decoding, where at each decoding step, the decoder receives $W$ rounds of syndrome measurements \cite{dennis2002topological}.
Among the corrections suggested by the decoder, only those corresponding to errors in the first $\lfloor\frac{W}{2}\rfloor$ rounds of the decoding window are applied.
In the next decoding step, the window is updated and advanced to include future syndrome information, while the rounds for which corrections have already been applied are removed from the window. 
If the window size $W$ is constant, data qubit errors with weight below $\lfloor (d-1)/2 \rfloor$ can be confused with measurement errors, causing long-lived error chains across windows \cite{huang2021constructions,kang2025quits}.
For toric codes with local checks, one round of syndrome extraction per window is insufficient for fault tolerance \cite{lin2024single,campbell2019theory}, while $O(d)$ rounds suffice \cite{dennis2002topological,huang2021constructions}.

Single-shot error correction provides an alternative \cite{PhysRevX.5.031043}. 
In single-shot error correction, after one round of syndrome extraction, a decoder is applied directly to the noisy syndrome to determine and implement corrections. 
To enable this, the syndrome must be obtained from a set of single-shot checks. 
It has been shown that such a check set exists for any stabilizer code \cite{campbell2019theory}. 
Specifically, one can always construct a set of checks ${G_i}$ and a corresponding set of single-qubit errors ${E_i}$ such that each $G_i$ anticommutes with a distinct $E_i$. 
A measurement error on $G_i$ can then be interpreted by the decoder as a data qubit error $E_i$, so that after applying the inferred correction, the residual error weight is upper bounded by the number of measurement errors.
Based on this principle, a single-shot check set for the toric code has been shown to achieve a sustainable threshold of $5.6\%$ under a phenomenological noise model with equal data and measurement error rates. 
However, the checks in this construction have unbounded weight as the code distance increases, which prevents a threshold under circuit-level noise where measurement error rates can scale with gate depth \cite{lin2024single}.

The above discussion assumes syndrome extraction is performed using unencoded syndrome qubits, or Shor-style gadgets \cite{96shor,divincenzo1996fault}, where each stabilizer is measured using a verified cat state. 
If alternative syndrome extraction gadgets are allowed, the number of repeated measurement rounds can be reduced to $O(1)$ \cite{Zheng_2020} by using Steane-style \cite{steane1997} or Knill-style \cite{knill2005quantum} syndrome extraction, where the syndrome qubits are encoded into logical $|0\rangle_L$ or $|+\rangle_L$ states of the same code. 
Ref.~\cite{huang2021between} and Ref.~\cite{huang2021constructions} generalize this approach by constructing syndrome extraction gadgets of various sizes and introducing the notion of the time distance $d_t(t_1,t_2)$ in the decoding graph, defined as the length of the shortest error path connecting time slices $t_1$ and $t_2$. 
They show that a decoding window spanning syndrome extraction rounds $t_i$ to $t_i+W-1$ is fault-tolerant if $d_t(t_i,t_i+W)=\Omega(d)$.   
Here, $t_i+W$ labels the round immediately following the window that is used as the time boundary in the time distance calculation. 
Under this condition, all errors of weight less than $\lfloor \frac{d-1}{2}\rfloor$ occurring in the first half of the decoding window are fully corrected before the next window is processed.
Using syndrome gadgets of size $l \times l$ arranged with appropriate temporal offsets, they reduce the required number of measurement rounds per decoding window to $O(d/l)$ due to an increased time distance. 
However, these approaches require preparing encoded syndrome states with high fidelity, which introduces additional overhead from post-selection \cite{reichardt2004improved} or verification steps \cite{Divincenzo2007slow}.

In what follows, we modify single-shot checks using a similar time-distance perspective to construct local single-shot checks with bounded support and use a dynamic measurement schedule to reduce the number of syndrome extraction rounds per decoding window without requiring encoded syndrome qubits.

\section{Check constructions}\label{sec: checks}
We begin by introducing the single-shot checks for the toric code, and then describe how to impose locality constraints to obtain bounded-weight checks. 
For clarity, we focus on $Z$-type checks; the $X$-type checks can be constructed analogously.

\subsection{Single-shot checks}
\begin{figure*}[htbp!]
    \centering
    \includegraphics[width=0.95\linewidth]{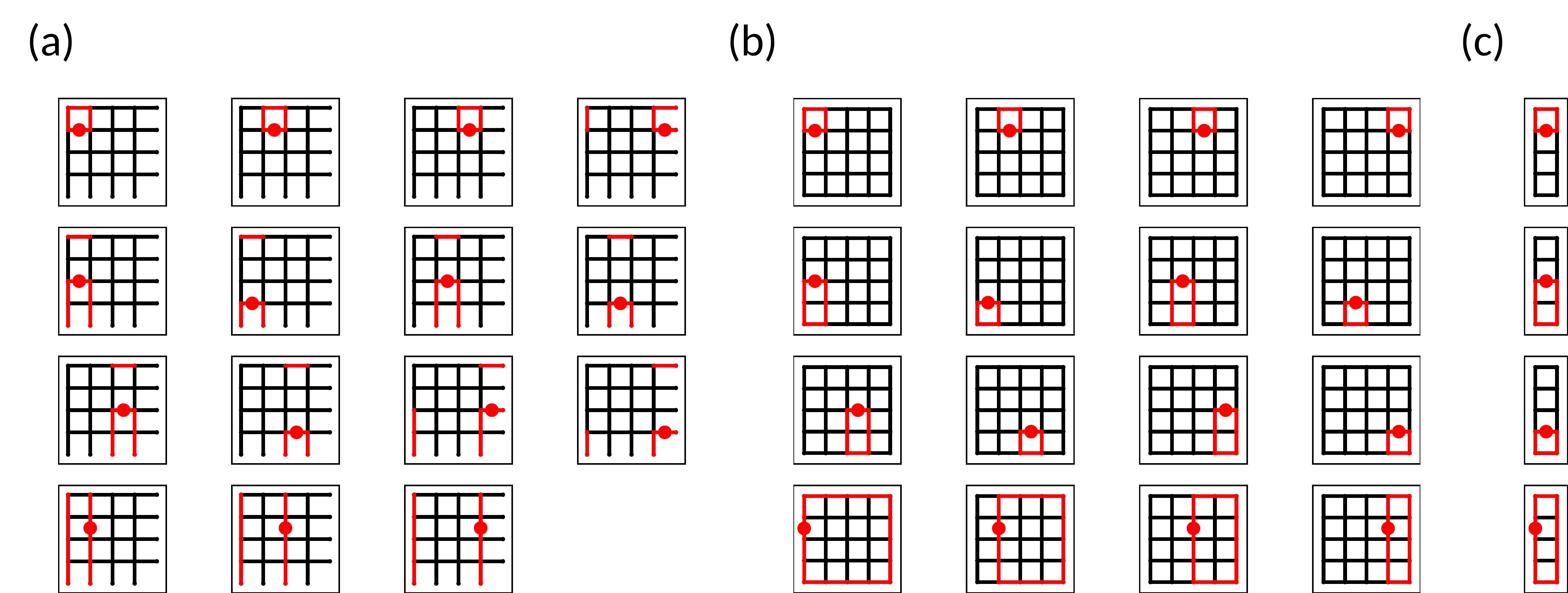}
    \caption{Illustration of different checks. In this representation, each link represents a qubit. We only show $Z$-type checks as $X$-type checks can be similarly obtained. The red lines in each subplot mark the qubits in support of each check. (a) Single-shot checks for an $L=4$ toric code. There are $L^2-1$ single-shot checks in total. Within each check, we use a red dot to mark the data qubit onto which we map the measurement error of the corresponding single-shot check. (b) Variable-width checks defined on a $l=4$ patch. (c) Fixed-width checks defined on a $l=4$ patch. These are two choices of local single-shot checks defined by different partitions of the torus. We use a red dot to mark the data qubits that are checked only by one variable-width check within the patch. If the data qubit is on the boundary of the patch, we map the measurement error of the corresponding local single-shot checks onto a time edge. Otherwise, we map the measurement error to a data qubit error on that data qubit instead. Note that the fixed-width checks all have the same width $1$, while the variable-width checks can have various widths.}
    \label{fig:single_shot_checks}
\end{figure*}

Single-shot checks for toric code were first introduced in Ref.~\cite{lin2024single}.
They can be constructed by performing Gaussian elimination on the Z(X)-type parity check matrices. 
The single-shot checks are not unique.
We can always reorder the qubits in the parity check matrix and obtain a different set of single-shot checks that map measurement errors onto a different set of qubits. 
Here, we adopt a slightly different set of single-shot checks than those used in Ref.~\cite{lin2024single}, as this modified form makes the subsequent construction of local single-shot checks straightforward. 
Figure~\ref{fig:single_shot_checks}(a) illustrates the $Z$-type single-shot check set used in this work. 
The check set consists of three families:

\begin{itemize}
\item[(a)] Square checks $S_i=\Box_{0,i}$, where $i=0,\ldots,L-1$.

\item[(b)] Rectangular checks
$R_{i,j} = \prod_{k=i}^{L-1} \Box_{k,j}$, where $i = 2, \ldots, L-1$ and $j = 0, \ldots, L-1$.

\item[(c)] Circular checks $C_i = \prod_{m=0}^{i}\prod_{n=0}^{L-1} \Box_{n,m}$, where $i = 0, \ldots, L-2$. 
\end{itemize}

Here, $\Box_{i,j}$ represents a local $Z$-type stabilizer located at plaquette $(i,j)$.
Each single-shot check is associated with a unique data qubit such that a measurement error on the check can be interpreted as a single-qubit data error on that qubit.
Note that there is no redundant check in the single-shot check set. 
Such a check set involves checks whose weight increases with the code distances. 
Therefore, they do not exhibit a threshold in an error model where the measurement error rate scales as the weight of checks.

\subsection{Local single-shot checks}
\begin{figure*}[htbp!]
    \centering
    \includegraphics[width=0.7\linewidth]{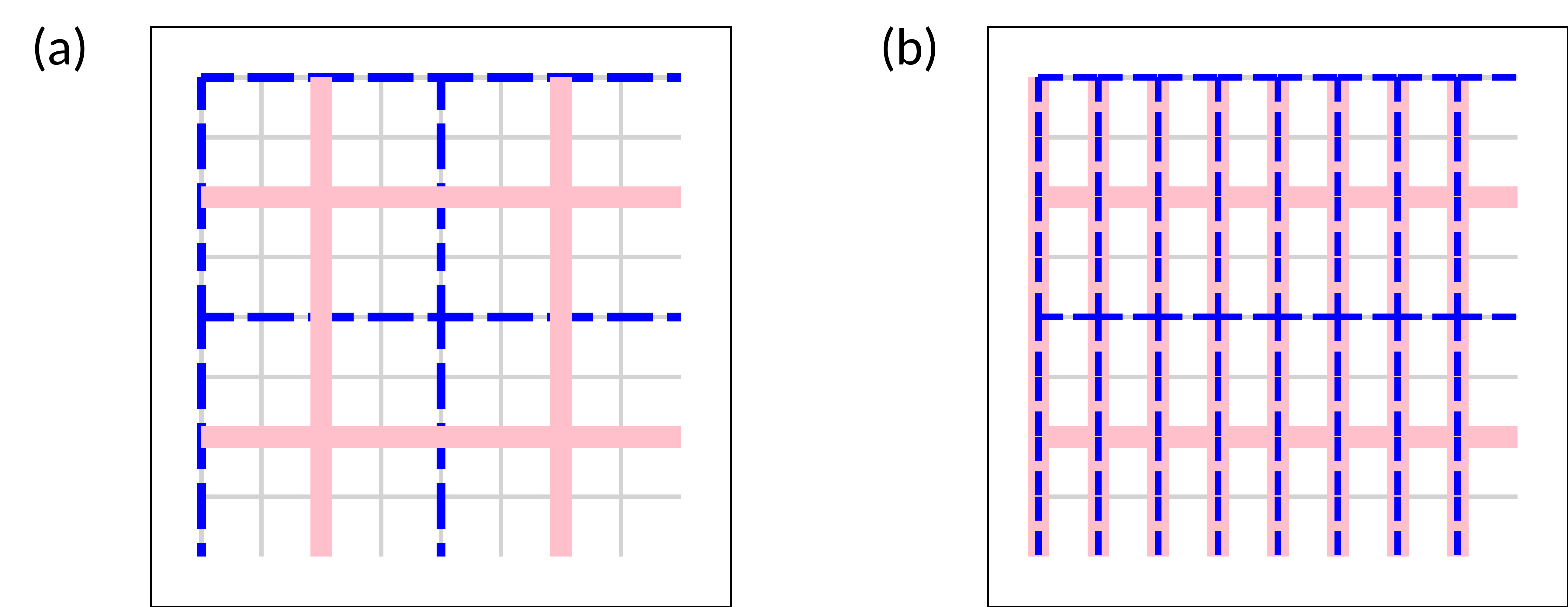}
    \caption{Illustration of the alternating partitions. (a) The alternating square partition used for the variable-width checks. (b) The alternating rectangular partition used for the fixed-width checks. The gray grids represent the torus, with links representing data qubits. The blue and pink lines mark the partition used for odd syndrome extraction rounds and even syndrome extraction rounds respectively. }
    \label{fig:partitions}
\end{figure*}

To avoid unbounded check weights, we introduce local single-shot checks. 
The idea is to partition the lattice into small $l\times l$ patches and define single-shot checks within each patch.
The blue lines in Fig~\ref{fig:partitions}(a) show an example of such a partition for $l=4$ on an $8\times 8$ lattice. 
For each patch, there are three types of checks defined on the patch that starts with $\Box_{u,v}$:
\begin{itemize}
\item[(a)] Square checks $S^{u,v}_i=\Box_{u,i+v}$, where $i=0,\ldots,l-1$.

\item[(b)] Narrow rectangular checks
$N^{u,v}_{i,j} = \prod_{k=i}^{l-1} \Box_{k+u,j+v}$, where $i = 2, \ldots, l-1$ and $j = 0, \ldots, l-1$.

\item[(c)] Wide rectangular checks $W_i^{u,v} = \prod_{m=i}^{l-1}\prod_{n=0}^{l-1} \Box_{u+n,v+m}$, where $i = 0, \ldots, l-1$. 
\end{itemize}

For the rest of the paper, we refer to such a set of checks as the \emph{variable-width} local single-shot checks. 
In total, there are $L^2$ variable-width checks. 
When $l=L$ and periodic boundary conditions are applied, $W^{u,v}_0$ becomes trivial and the remaining checks reduce to the full single-shot check set.
If $l\le2$, we no longer have the narrow rectangular checks $\{N_{i,j}^{u,v}\}$. If we further reduce $l$ to $l=1$, the wide rectangular checks $\{W^{u,v}_{i}\}$ do not exist either, and the check set returns to the local check set. 
Figure~\ref{fig:single_shot_checks}(b) shows the variable-width checks defined for a $4\times 4$ patch. 

Similar to single-shot checks, each check within a patch is associated with a data qubit not shared by the other checks in the same patch. 
This means that for most checks, measurement errors are thus mapped to data errors on these qubits. 
The exceptions are the checks $W_0^{u,v}$ of all patches (i.e. the down left check in Fig.~\ref{fig:single_shot_checks}(b)). 
For these checks, their corresponding data qubits are on the boundary of the patches. 
Within the same patch, this data qubit is not checked by other checks. 
However, they are checked by the checks defined in the neighbor patch. 
To handle the measurement errors on these checks, we introduce vertical time edges in the decoding graph that connect their corresponding detectors at the same position on neighboring time layers. 
The details of constructing the decoding graph will be discussed in Sec~\ref{sec: decoder}.

Due to the existence of the wide rectangular checks, the variable-width checks can induce hook errors when being implemented at the circuit level \cite{dennis2002topological}. Hook errors are correlated errors propagated from syndrome qubits onto the data qubits through CNOT gates.
Such error mechanism necessitates either encoded syndrome qubits or a careful scheduling of the entangling gates.
To avoid such complications, we introduce a second set of checks which we refer to as the \emph{fixed-width} local single-shot checks. 
To define fixed-width checks, we instead partition the lattice into $l\times1$ strips. 
An example of $l=4$ is shown in blue lines on an $8\times 8$ lattice in Fig.~\ref{fig:partitions}(b). 
For each strip that starts with $\Box_{u,v}$, we similarly define checks within the strip as:

\begin{itemize}
\item[(a)] Top-boundary checks $\tilde{T}^{u,v}=\Box_{u,v}$.
\item[(b)] Bottom-boundary checks
$\tilde{B}^{u,v}_{i} = \prod_{k=i}^{l-1} \Box_{k+u,v}$, where $i = 2, \ldots, l-1$.
\item[(c)] Full-patch checks $\tilde{F}^{u,v}=\prod_{i=0}^{l}\Box_{u+i,v}^{u,v}$.
\end{itemize}

We will use boundary checks to refer to both $\{\tilde{T}^{u,v}\}$ and $\{\tilde{B}^{u,v}_{i} \}$. 
If $l\le2$, the bottom-boundary checks do not exist. When $l=1$, the top-boundary checks and the full-patch checks overlap and we again reach the local checks. 
Figure.~\ref{fig:single_shot_checks}(c) shows an example of fixed-width checks defined on a $4\times 1$ strip. 
Similar to the variable-width checks, we can find a data qubit for each check that is not checked by other checks belonging to the same strip.
If these data qubits are not on the boundary of the strip, we map measurement errors on their corresponding checks onto errors on these qubits, which then become space edges in the decoding graph; otherwise, we map them onto a vertical time edge in the decoding graph. 
Due to the fixed-width structures of the checks, we can find a syndrome extraction circuit where hook errors are harmless. 
The circuit construction is detailed in Sec.~\ref{sec:circuit}.

We now describe how these local single-shot checks allow us to reduce the number of syndrome extraction rounds.

\subsection{Dynamic checks}{\label{sec:dynamic_checks}}

\begin{figure*}[htbp!]
    \centering
    \includegraphics[width=0.95\linewidth]{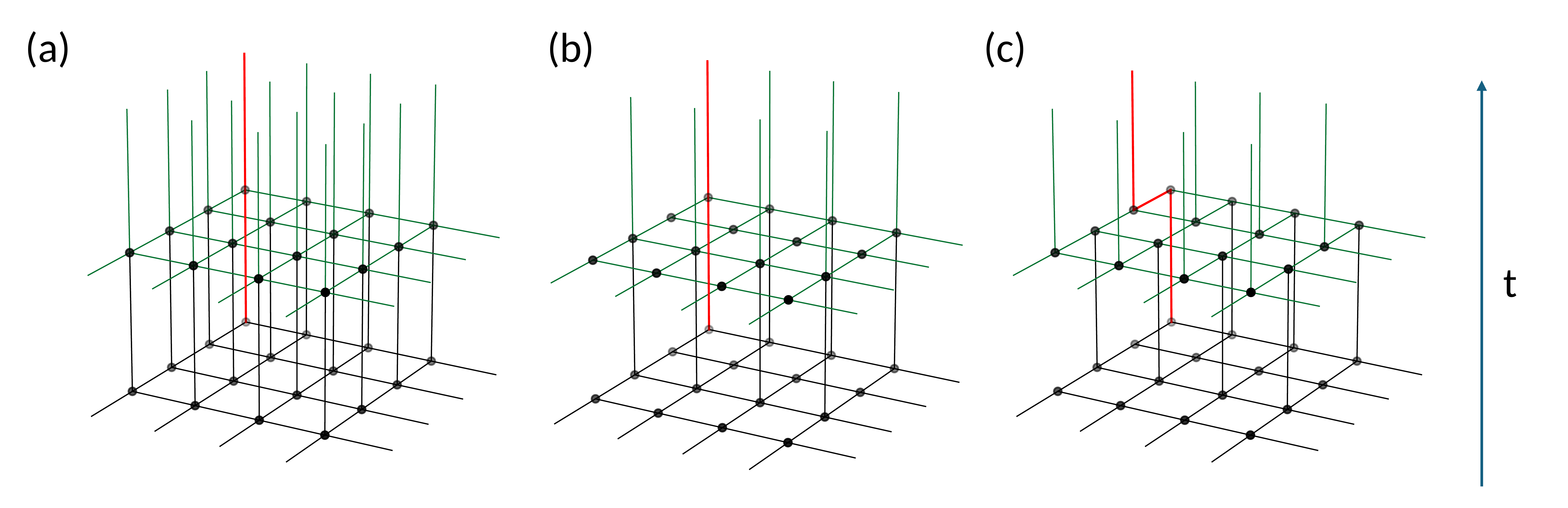}
    \caption{Illustration of the decoding graph used for different measurement schemes under a phenomenological noise model. Each plot shows two consecutive time layers of the decoding graph: an even layer and an odd layer. Black lines denote the detectors within the first layer and the time edges connecting them to the second layer. Green lines similarly denote the detectors within the second layer and the time edges connecting them to future layers. Each black dot represents a detector. Edges within each layer are space edges representing a data qubit error, or effective data qubit errors resulting from a mapped measurement error. Vertical edges connecting the layers are time edges representing measurement errors. The red line shows the shortest path to go from a detector in the first layer to the time boundary of the given decoding volume. The length of the path gives the time distance $d_t$ of the decoding volume. (a) Decoding graph for local checks. Each detector is defined by the product of consecutive measurements of the same check. Within each layer, the space edges form a torus. Across layers, we have time edges between each pair of consecutive detectors located at the same spatial position in different layers. The length of the shortest path connecting to the time boundary equals the number of layers. (b) Decoding graph for the fixed-width checks defined on $l=2$ patches with an aligned measurement scheme. Here, the detectors are constructed by first converting the local single-shot check syndrome to the local check syndrome and then taking the products of the consecutive local check syndrome. In the decoding graph, the space edges have the same structure as the local checks. The time edges are sparser and only occur in certain rows of each layer. However, the length of the shortest path to the boundary is still equal to the number of layers as the time edges are aligned. (c) Decoding graph for the fixed-width checks defined on $l=2$ patches with an offset measurement scheme. The detectors are defined as in the previous case. The time edges of each layer do not align in this case, but they can only be located at certain rows, depending on whether the time layer is odd or even. This offset alignment of time edges increases the length of the shortest path to the time boundary in the decoding graph, which is why the optimal decoding window can contain fewer syndrome extraction rounds than the local checks.}
    \label{fig:decoding_graph}
\end{figure*}

Before introducing the details of our construction, we revisit the conventional measurement scheme from the perspective of time distance.
For local checks, the decoding graph has the same spatial structure in every layer, and the time edges align across layers.
An example of the decoding graph for local checks with window size $W=2$ is shown in Fig.~\ref{fig:decoding_graph} (a). 
Because the shortest path connecting the first time layer to the final time boundary requires traversing one vertical edge per layer, we have $d_t=W$ for the local checks within a single decoding window. 
Thus, we need at least $W=O(d)$ rounds of syndrome extraction for each decoding window to achieve fault tolerance when using local checks.

To reduce the number of required measurement rounds, it is therefore not sufficient to repeatedly measure the local single-shot checks from Sec.~\ref{sec: checks}.
The reason can be seen from the fact that for each patch, we map the measurement errors of one of the checks onto a time edge.
These time edges form a path connecting through the whole decoding window, as shown by the red edges in Fig~\ref{fig:decoding_graph}(b). 
Similar to the local checks, we have $d_t(t_i,t_i+W)=W$, which implies that $O(d)$ rounds of syndrome extraction are needed to achieve $d_t(t_i,t_i+W)=d$, as is required by the theorem.

In order to reduce measurement, we introduce a dynamic measurement scheme for the local single-shot checks.
Specifically, in each round of syndrome extraction, we measure the local single-shot checks defined on patches belonging to different partitions of the torus.
For example, Fig.~\ref{fig:partitions}(a) shows the alternating partitions of the torus we use to define the dynamic measurement scheme of the variable-width checks. 
In all odd measurement rounds, we measure the checks associated with the blue partition; in even rounds, we measure those associated with the pink partition.
These two partitions are related by a translation of $\lfloor l/2\rfloor$ in both directions.
Similarly, for the fixed-width checks, we alternate between the two partitions (shown by blue and pink lines in Fig.~\ref{fig:partitions}(b)) and measure the fixed-width checks defined on these patches in alternating rounds. 
In this case, the two partitions are related by a downward translation of $\lfloor l/2\rfloor$.

Under this dynamic measurement scheme, the time edges in the decoding graph become offset.
An example of the decoding graph for the fixed-width checks with $l=2$ on a $4\times4$ torus is shown in Fig.~\ref{fig:decoding_graph} (c). 
In this case, the shortest error path connecting the first layer to the time boundary must traverse both time edges and space edges.
The red line in Fig.~\ref{fig:decoding_graph}(c) illustrates such a shortest path.
Because of the offset, we now have, $d_t(t_{i},t_{i}+W)>W$, implying that fewer than $d$ rounds of syndrome extraction are sufficient to achieve $d_t(t_i,t_{i}+W) = d$. 

The amount by which the number of syndrome extraction rounds can be reduced depends on $l$. 
This follows from the fact that the shortest error path must cross a number of space edges proportional to the translation distance of the partitions, which scales as $O(l)$ for both types of local single-shot checks. 
For example, in the case $l=2$, in fixed-width checks, we translate the partition downward by $\lfloor l/2\rfloor=1$ in every syndrome extraction round. 
Fig.~\ref{fig:decoding_graph} shows that the shortest path to the boundary crosses exactly one space edge on the second layer.

Based on this observation, for the fixed-width checks with a dynamic measurement scheme, the relation of the time distance and the number of syndrome extraction rounds is given by:
\begin{equation}
    d_t(t_i,t_{i}+W)=W+(W-1)\lfloor\frac{l}{2}\rfloor
\end{equation}
As we can verify, in Fig~\ref{fig:decoding_graph}(c) we have two layers, and the shortest path has a length of $3$. 
To achieve $d_t=\Omega(d)$, we therefore require:
\begin{equation}\label{eq:optimal}
    W\ge \frac{d+\lfloor l/2\rfloor}{1+\lfloor l/2\rfloor}
\end{equation}

Similarly, for variable-width checks, the time distance of a decoding window size $W$ is given by:
\begin{equation}
    d_t(t_i,t_{i}+W)=W+2(W-1)\lfloor \frac{l}{2}\rfloor
\end{equation}
The required decoding window size is given by:
\begin{equation}
    W\ge \frac{d+2\lfloor l/2\rfloor}{1+2\lfloor l/2\rfloor}
\end{equation}
Note that when $l=d$, the check $W_0^{u,v}$ becomes an identity and is thus removed from the check set. Therefore, no time edge is added to the decoding graph and $d_t=\infty$, which leads to the requirement $W\ge 1$. 

We will describe the details of the decoder construction in Sec.~\ref{sec: decoder}. 
Before that, in Sec.~\ref{sec:circuit}, we explain how to implement the local single-shot checks at the circuit level.

\section{Syndrome extraction circuit}\label{sec:circuit}
We discuss the construction of the syndrome extraction circuits for the local single-shot checks below.
To reliably extract stabilizer measurement outcomes in the presence of circuit-level noise, we use a syndrome extraction circuit (SEC) that enforces fault tolerance.
Each stabilizer measurement employs a single syndrome qubit that interacts with the data qubits in the stabilizer support through a sequence of CNOT gates.
The ordering of these CNOTs is critical: a single fault on the syndrome qubit can propagate to multiple data qubits, producing hook errors.
If not properly controlled, such hook errors can create low-weight correlated data errors that reduce the effective distance of the code.

\begin{figure}[htbp!]
    \centering
    \includegraphics[width=0.99\linewidth]{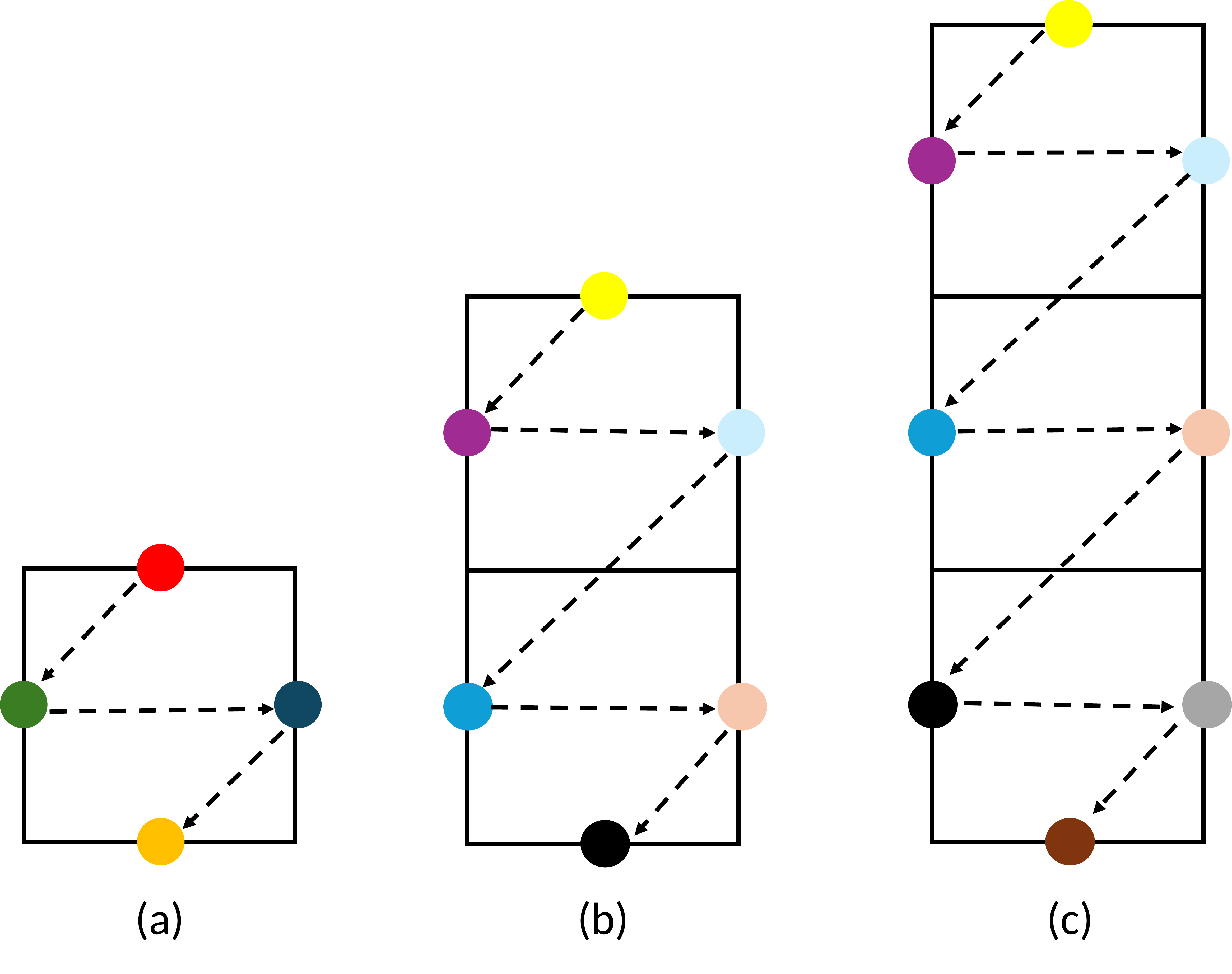}
    \caption{Examples of syndrome extraction CNOT orderings for the local single-shot checks. Each colored data qubit indicates the order in which the ancilla interacts with the data via CNOT gates. (a) Square check. (b–c) Narrow rectangular checks, where the zigzag pattern prevents hook-error propagation.}
    \label{fig:SECs}
\end{figure}

The local single-shot checks we consider have either the same square geometry as the original local checks, or they appear as narrow rectangular checks, as illustrated in Fig.~\ref{fig:single_shot_checks} and Fig.~\ref{fig:SECs}.
We focus on checks of these shapes because a circuit using a bare syndrome qubit can be constructed for them in a way that ensures any resulting hook errors are harmless.

With these checks in mind, we now describe the corresponding circuit construction.
The CNOT directions in the circuit follow the NEWS convention \cite{kang2025quits,PhysRevLett.129.050504}, and the gate scheduling avoids conflicts between adjacent stabilizers, thereby preventing unintended entanglement between syndrome qubits. 
We further adopt zigzag CNOT patterns for the narrow rectangular checks, as illustrated in Fig.~\ref{fig:SECs}. 
Figure~\ref{fig:SECs}(a) shows the ordering for a square check, while Fig.~\ref{fig:SECs}(b) and (c) show the corresponding orderings for narrow rectangular checks. 
The colored data qubits indicate the time ordering of the ancilla–data CNOT interactions. 
In our implementation, the circuits for all square checks are applied first, followed by those for the rectangular checks. 
Finally, we note that all $X$-type and $Z$-type checks are extracted in parallel.

For both the square and rectangular checks, this zigzag pattern ensures that error propagation aligns with directions in which hook errors remain harmless, either equivalent to a single data-qubit error or mapped to a benign direction with respect to the logical operators. 
Therefore, these hook errors do not reduce the effective circuit distance of the code, which can be defined as the minimum number of error mechanisms that can cause an undetectable logical error.
Similar ordering strategies have been used to maintain circuit-level fault tolerance in other codes~\cite{PhysRevA.98.050301,PhysRevA.101.042312,PhysRevA.90.062320}.
We note, however, that this ordering does not always guarantee the existence of a threshold.

\section{Noise model}\label{sec:noise_model}
We consider two different noise models: a phenomenological noise model and a circuit-level noise model.
We describe each in detail below.

\subsection{Phenomenological model}{\label{sec:phenom}}

Phenomenological noise models are commonly used to analytically study single-shot error correction properties \cite{campbell2019theory,PRXQuantum.2.020340} and to numerically evaluate the performance of single-shot error correction codes \cite{higgott2023improved,breuckmann2021single,PRXQuantum.2.020340,kubica2022single,brown2016fault}. 
This model simplifies error analysis by only accounting for data-qubit and measurement errors.
During each round of syndrome extraction, we assume these errors are independent and identically distributed. 
Since we focus on $Z$-type checks, we consider each data qubit to experience a bit-flip channel $C(\rho)=(1-p)\rho +pX\rho X$, where $X$-type errors occur with probability $p$ on each data qubit. 
In addition, the measurement outcome of each check is flipped with probability $q$, and in this work we take $p=q$.

\subsection{Circuit-level noise model}{\label{sec: circuit-noise}}
In circuit-level implementation, additional error mechanisms can degrade the code performance, especially when reduced measurement schemes are considered.
While the effect of hook errors can be suppressed by the circuits described in Sec.~\ref{sec:circuit}, internal errors remain a challenge \cite{delfosse2021beyond}.
Internal errors are errors on data qubits, that occur during one round of syndrome extraction after some of checks have been measured. These errors can originate from gate errors and hook error propagation.
They evade the detection of the already completed checks, but are detectable by subsequent ones.
This results in inconsistent syndrome information within a single round, complicating the decoding and potentially compromising the single-shot property of the code. 

In this work, we consider the following circuit-level noise model.
Before each syndrome extraction round, every data qubit is subject to single-qubit depolarizing noise with probability $p$. 
Each CNOT gate is followed by two-qubit depolarizing noise with probability $p$ acting on the participating qubits, and each H gate is followed by single-qubit depolarizing noise with probability $p$. 
We neglect idling errors to avoid complicated circuit parallelization and optimization. 
Finally, both qubit reset and measurement operations are assumed to undergo depolarizing errors with probability $p$.

As discussed in Sec.~\ref{sec: decoder}, internal errors from this error model can affect the structure of the decoding graph, increasing the complexity of hyperedge decomposition and reducing the effective distance of the code. Importantly, in our circuit construction, a single error mechanism can contribute at most a weight-2 internal error on the data qubits. 
This ensures that internal errors remain decodable.

\section{Decoders}\label{sec: decoder}

Typically, toric codes with local checks can be decoded with an MWPM decoder, as every fault flips two detectors when $Y$ errors are decomposed into products of independent $X$ and $Z$ errors, allowing $X$-type checks and $Z$-type check syndromes to be decoded independently. 
For the local single-shot checks constructed in this work, this property does not hold directly. 
In order to apply MWPM decoders, we need to perform hyperedge decomposition. 
Here we describe how the decoding graphs are constructed for both noise models, as this construction is central to our decoding procedures.

For the local single-shot checks, instead of constructing detectors directly by taking products of syndromes from consecutive syndrome extraction rounds, we first transform the syndrome of the local single-shot checks into the syndrome of the local checks.
This follows from the fact that the parity-check matrix of the local checks, $P$, and the parity-check matrix of the local single-shot checks, $\tilde{P}$, are related by an invertible matrix $U$, such that $P = U\tilde{P}$.
Accordingly, the syndrome of the local checks $M(t)$ can be obtained from the syndrome of the local single-shot checks $\tilde{M}(t)$ by $M(t)=U\tilde{M}(t)$. 
Once we obtain the local-check syndrome, the detectors are constructed in the same way as for the standard local checks.
Based on this detector conversion, error mechanisms in the dynamic single-shot syndrome extraction circuits can be mapped to combinations of data-qubit errors and measurement errors on local checks.

We now describe the three types of edges, time edges, space edges, and space-time edges, in the decoding graph.
Time edges are vertical edges that connect detectors at the same spatial position in rounds $t$ and $t+1$.
We can interpret them as measurement errors on corresponding local checks. 
Space edges connect detectors at the same time slice $t$. 
These are the edges associated with data qubit errors. 
Finally, space–time edges refer to edges connecting detectors at different spatial positions belonging to neighboring time layers. 
These edges can be interpreted as combinations of data qubit errors and measurement errors on local checks.

\begin{figure*}[htbp!]
    \centering
    \includegraphics[width=0.7\linewidth]{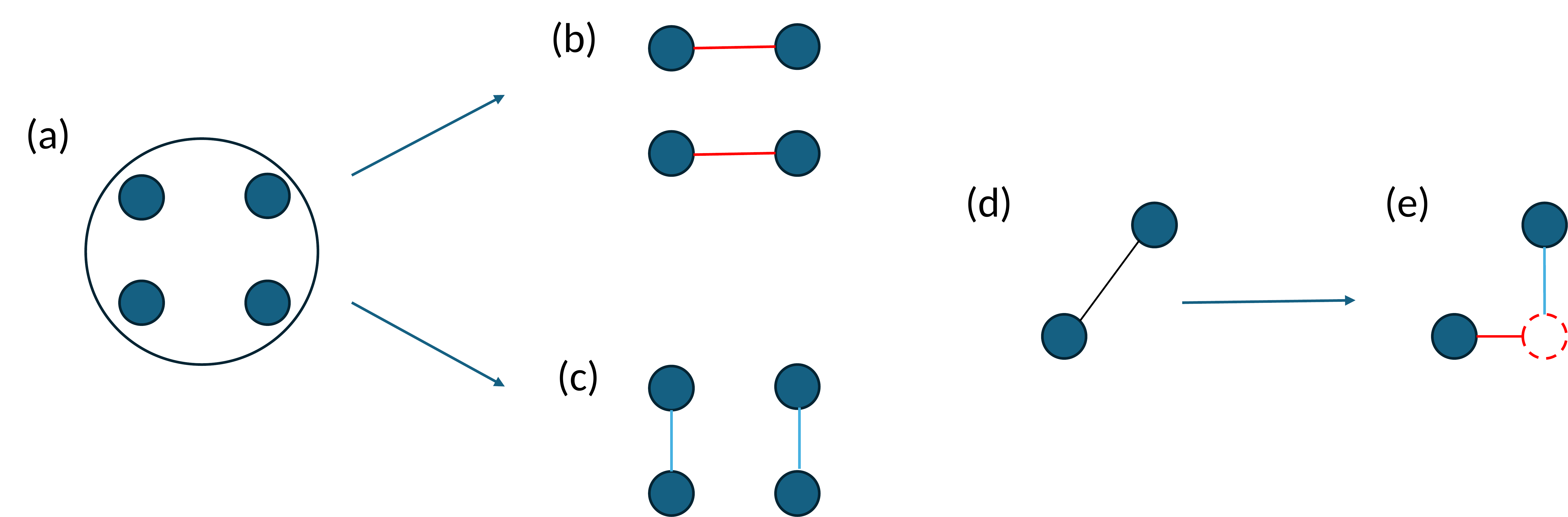}
    \caption{Illustration of different types of hyperedge decomposition. The vertical direction represents time, and the horizontal direction represents space. (a) A hyperedge (represented by a circle) connecting four detectors (blue dots). The detectors are located on two neighboring time layers (represented by two rows). A measurement error on any local single-shot check can cause this hyperedge, unless the check spans all boundary qubits of the patch. Hyperedges cannot be directly processed by MWPM algorithms and therefore need decomposition. (b) Space-edge-first decomposition: The hyperedge is decomposed into two space edges (red lines) on neighboring time layers, which corresponds to two data qubit errors happening at consecutive rounds of syndrome extraction. This decomposition is also used for the phenomenological noise model. (c) Time-edge-first decomposition: The hyperedge is decomposed into two time edges (blue lines). This effectively maps a measurement error on the local single-shot checks to two measurement errors of local checks. (d) A space-time edge representing a circuit error mechanism whose starting or ending detector is not contained within the allowed sets defined by the phenomenological decoding graph. Such edges are typically caused by an internal error. When using the time-edge-first decomposition, we directly add such an edge into the decoding graph. (e) In space-edge-first decomposition, since the original space-time edge in (d) is not allowed, we introduce an " auxiliary" detector (red dashed-line circle) by projecting the detector on time layer $t+1$ to time layer $t$. The original space-time edge is decomposed into a time edge connecting the detector in layer $t+1$ and the auxiliary detector, and a space edge connecting the auxiliary detector and the other detector on layer $t$.  }
    \label{fig:decomposition_meas}
\end{figure*}

\subsection{Decoder for the phenomenological noise models}
In the phenomenological noise model, we only need to handle data qubit errors and measurement errors. 
In this case, the data qubits errors occurring at syndrome extraction round $t$ can be mapped to space edges in the decoding graph at time layer $t$. 
Measurement errors produce two different structures. 
Specifically, measurement errors on the checks that have all their support on the patch boundaries (i.e. the full-patch checks of the fixed-width checks or the largest wide rectangular checks of the variable-width checks) are mapped to time edges connecting detectors in adjacent layer, as measurement errors on these checks will be mapped onto measurement errors on one of the local checks involved in the same patch.
For fixed-width checks, such time edges are located at the second row of the local check detectors in each patch. 
For variable-width checks, the time edges are located only at the local check detectors at position $(u+1,v)$ of each patch. 
All other measurement errors form degree-4 hyperedges that involve the same pair of detectors in different time layers. 
An example of these degree-4 hyperedges is given in Fig.~\ref{fig:decomposition_meas}(a). 
These errors can be interpreted as two errors occurring at neighboring time slices on the same data qubits. 
Based on this interpretation, we can always decompose these hyperedges into two edges corresponding to the data qubit errors, as is shown in Fig.~\ref{fig:decomposition_meas}(b). 
Following this, we obtain decoding graphs for both aligned and offset measurement schemes, with examples shown in Fig.~\ref{fig:decoding_graph} (b) and (c).

Note that the time edges in the decoding graph can only occur at certain positions depending on whether they start from odd or even time layers.
We call these positions allowed positions for time edges.
If we have a time edge that goes from position $(i,j,t)$ to $(i,j,t+1)$, we call $(i,j,t)$ the allowed starting points and $(i,j,t+1)$ the allowed ending points of a time edge. 
We will use these positions to define the decoding graphs for the circuit-level model.

\subsection{Decoder for a circuit-level noise model}{\label{sec:circuit-decoder}}
 
The additional error mechanisms at the circuit level, especially the internal errors, lead to various hyperedge structures and pose a challenge to hyperedge decomposition. 

Here we focus only on the error mechanisms in the circuit we design for the fixed-width local single-shot checks with $l\le 4$. 
We describe the $Z$-type decoding graph, and the $X$-type case follows analogously.
In this case, the allowed starting positions and ending positions of time edges occur in rows. 
The degree of hyperedges from the circuit-level noise model can only be either $4$ or $6$.
The degree-$4$ hyperedges can be further divided into two types. 
The first type involves a pair of detectors that have the same spatial positions on neighboring time layers, which are typically caused by measurement errors. 
The second type only has one such pair. 
The remaining two detectors can be on the same or different time layers, with different space positions. 
This type of hyperedges is mainly introduced by internal errors.
Each degree-$6$ hyperedge involves two pairs of detectors that share the same spatial positions but on neighboring time layers and a remaining pair of detectors that are on the same time layer but different spatial positions.
Such hyperedges are also introduced by internal errors.

Different ways of handling the hyperedges give different decoding performances and change code properties. 
Here, we introduce two different decompositions that produce different decoding graphs, namely, the time-edge-first decomposition and the space-edge-first decomposition. 
Note that we require all edges we add into the decoding graph to remain spatially local in the sense that their projection on the same time layer are either nearest or second nearest neighbors. 
Spatially, they correspond to either a data qubit error or two data qubit errors occurring on one vertical link and one horizontal link of the lattice that can be caused by a hook error.
This constraint is to ensure that the number of edges needed to form a logical loop is always lower bounded by the code distance $d$.

In the time-edge-first decomposition, we add all the edges as usual. 
For hyperedges, we prioritize adding time edges. 
Specifically, among the detectors connected by the same hyperedge, as long as a pair of detectors with the same spatial location located on neighboring time layers occur, we add a time edge between these detectors. 
Due to the structure of the hyperedges, there can be at most a pair of detectors left after adding time edges, which are guaranteed to be spatially local. 
We then directly connect the remaining two detectors in the decoding graph. 
An example of decomposing a degree-$4$ hyperedge caused by a measurement error is shown in Fig.~\ref{fig:decomposition_meas}(c), where the hyperedege is decomposed into two time edges.
We show all other possible decompositions in time-edge-first decomposition in App.~\ref{app:time-edge}.  
Note that we only decompose hyperedges using the detectors that are in support of the hyperedges. 
Since we do not impose any constraints on adding time edges, such a decomposition does not increase the time distance across different time layers.
Therefore, it is expected that decoding with a graph constructed using time-edge-first decomposition cannot reduce syndrome extraction rounds.

In the space-edge-first decomposition, we restrict the allowed space-time edges and time edges so that they can only start from the allowed starting points $(i,j,t)$ in each layer $t$, and end at the allowed ending position $(i,j',t+1)$  of the same spatial row in the next layer $t+1$ where $j$ can be different from $j'$. 
With such a restriction, the time distance between layers with the same checks is the same as in the phenomenological noise model. 
These edges and all the space edges form the set of allowed edges in this decomposition.

With the constraints on the allowed edges, we now only add the edges that fall into this category directly. 
For edges not belonging to such a category (see Fig.~\ref{fig:decomposition_meas} for example), we introduce auxiliary detectors that are not part of the detectors supporting the edge so that we can decompose the edge into two allowed edges.
As shown in Fig.~\ref{fig:decomposition_meas}(d), we project the detector at time layer $t+1$ onto time layer $t$ into an auxiliary detector. 
We can then decompose the edge into a time edge and a space edge, each connecting the auxiliary detector with one of the detectors in its support. 
As for the hyperedges, we prioritize adding space edges rather than time edges. 
For example, for a degree-$4$ hyperedge caused by measurement error in Fig.~\ref{fig:decomposition_meas}(a), we decompose it into two space edges, which correspond to data qubit errors on the same qubit at neighboring time slices, as is shown in Fig.~\ref{fig:decomposition_meas}. 
More details of the decomposition of various hyperedge decompositions can be found in App.~\ref{app:space-edge}. 

Note that hyperedge decomposition always leads to a loss of correlation information. 
Furthermore, introducing auxiliary detectors reduces the correlation information in the decoding graph.
Consequently, the decoding performance of a $W=d$ window based on space-edge-first decomposition is suboptimal as it loses distance. 
Nevertheless, we preserve the increased time distance between time layers. 
Therefore, using a decoding graph obtained from space-edge-first decomposition, we do not need as many measurement rounds to reach the optimal performance within this decoding graph.

In short, under time-edge-first  decomposition, the syndrome rounds within each decoding window cannot be reduced. 
However, because the edges are unrestricted, time-edge-first decomposition can retain more correlation information between detector flips and is therefore expected to yield a lower logical error rate for a fixed length of syndrome history.
In contrast, under space-edge-first decomposition, optimal decoding performance can be achieved with a reduced number of measurement rounds in each decoding window, leading to lower computation latency.
In practice, one can choose an appropriate balance between these decompositions to trade off decoding accuracy against latency.

We want to emphasize that for all the hyperedge decompositions we perform, if the resulting edges already exist in the decoding graph, we assume that these edges can be flipped by two independent error mechanisms and update the weights of the edges accordingly \cite{higgott2022pymatching}.

Note that various hyperedge decoders also exist \cite{roffe_decoding_2020, beni2025tesseractdecoder}. 
Typically, hyperedge decoding is computationally more expensive. 
Therefore, in this work we choose to take advantage of the MWPM decoder. 
A similar strategy, in which a MWPM decoder based on hyperedge decomposition, has been used for Floquet codes \cite{delfosse2023splitting}.

We compare the decoding performance under both these different decompositions in Sec~\ref{sec: results}.

\begin{figure*}[htbp!]
    \centering
    \includegraphics[width=0.99\linewidth]{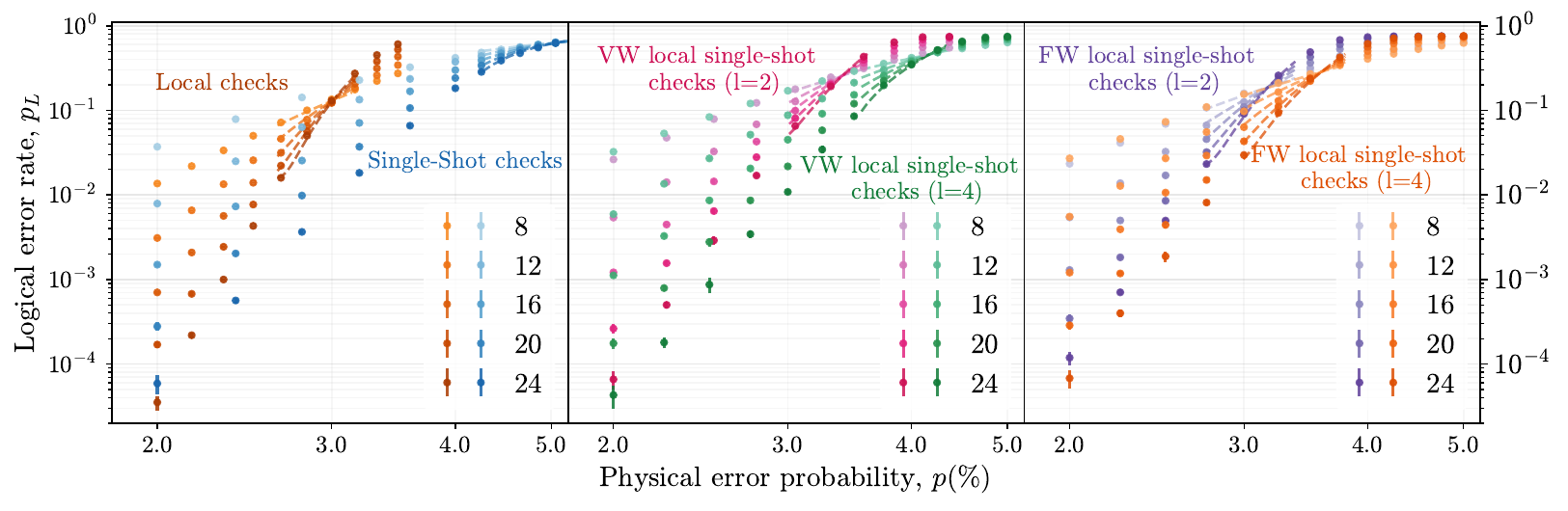}
    \caption{Error correction performance under a phenomenological noise model with $d+2$ rounds of syndrome extraction with full history decoding. Curves of different shades of the same color mark codes with different distances. (left) Comparison between local checks and single-shot checks. (center) Variable-width (VW) local single-shot checks defined on partitions of different $l$ using an offset measurement scheme. (right)  Fixed-width (FW) local single-shot checks defined on partitions of different $l$ using an offset measurement scheme. The legend indicates the code distance.}
    \label{fig:phenom}
\end{figure*}

\section{Numerical results}{\label{sec: results}}
We use the MWPM decoder from Pymatching \cite{higgott2022pymatching} for all simulations. 
All reported results in this section correspond to the $Z$-type logical error rate.
The $X$-type logical error rate can be similarly obtained due to the CSS property of the code. 
We collect up to $10^8$ samples for each data point.

We adopt the offset measurement scheme for local single-shot checks for all discussion in this section.
A comparison between offset and aligned measurement schemes for local single-shot checks can be found in App.~\ref{App:offse_vs_daligned}. 

\subsection{Phenomenological noise model}
We evaluate performance under the phenomenological noise model defined in Sec.~\ref{sec:phenom} and present the results below. 
For all the cases we investigate, we perform $d+2$ rounds of syndrome extraction to match the number of syndrome extraction rounds we use for the circuit-level simulation (See Sec~\ref{sec:circuit_result} for details). 
For all the checks, we construct decoding graphs for all $d+2$ rounds to decode rather than dividing the syndrome extraction rounds into decoding windows. 
For the single-shot checks, we ignore the redundant identity checks. In this way, we effectively still perform single-shot error correction as no time edge is added and the decoding graphs of different time layers are independent. 

The results of the different checks considered in this paper under the phenomenological noise model is plotted in Fig.~\ref{fig:phenom}. 
We compare performance in terms of both threshold and logical error rate scaling.
We fit the data close to the threshold using the function $p_l = ax+ bx^2 +c$, where $p_l$ is the logical error rate obtained from the simulation and $x=(p-p_{th})d^{1/\mu}$.
Here, $p$ is the physical error rate, $d$ is the code distance, $a,b,c$ and $\mu$ are fit parameters and $p_{th}$ is the threshold value obtained from the fit.

Among all the checks, local checks give the lowest threshold at $2.95\%$ as is shown by the yellow curves in Fig.~\ref{fig:phenom} (left). 
The single-shot checks shown by the purple curves in Fig.~\ref{fig:phenom} (left) give the highest threshold at $5.16\%$. 
Note that the single-shot threshold we obtain is slightly different from the sustainable threshold reported in Ref.~\cite{lin2024single} because the sustainable threshold is obtained by performing the same number of syndrome extraction rounds for codes of all distances.
However, the slopes of local check curves are larger than those of single-shot checks. 
We believe this reduction in slope results from the influence of residual data-qubit errors accumulated from previous rounds of error correction, as measurement errors are mapped onto data-qubit errors in single-shot error correction. 

Note that the direct comparison of the logical error rates between local checks and single-shot error correction shown in the plots is unfair.
In practice, when using single-shot error correction, only $O(1)$ rounds of syndrome extraction are required per logical operation. 
As a result, the same computation with single-shot error correction requires shorter runtime and thus a shorter syndrome history.
In our simulation, however, we use syndrome histories of the same length. 
For local checks, this corresponds to a single logical cycle, whereas for the single-shot checks, it corresponds to $O(d)$ logical cycles.
Nevertheless, for a consistent threshold comparison, we choose to report these logical error rates. 
The same argument applies to the logical error rates for the local single-shot checks with different $l$. 
A similar benchmarking methodology is used in Ref.~\cite{huang2021between} and Ref.~\cite{huang2021constructions} to compare the performance of syndrome qubits encoded in patches of different sizes.

The performance of the local single-shot checks under a dynamic measurement scheme falls between that of local checks and single-shot checks. 
For both the variable-width checks (shown in Fig.~\ref{fig:phenom} (center)) and fixed-width checks (shown in Fig.~\ref{fig:phenom} (right)), as $l$ increases, the threshold of the curves increases.
For variable-width checks, the threshold is $3.41\%$ for $l=2$, while the threshold is $4.16\%$ for $l=4$. 
For fixed-width checks, the threshold is $3.18\%$ for $l=2$ and $3.59\%$ for $l=4$. 

\begin{figure}[htbp!]
    \centering
    \includegraphics[width=0.9\linewidth]{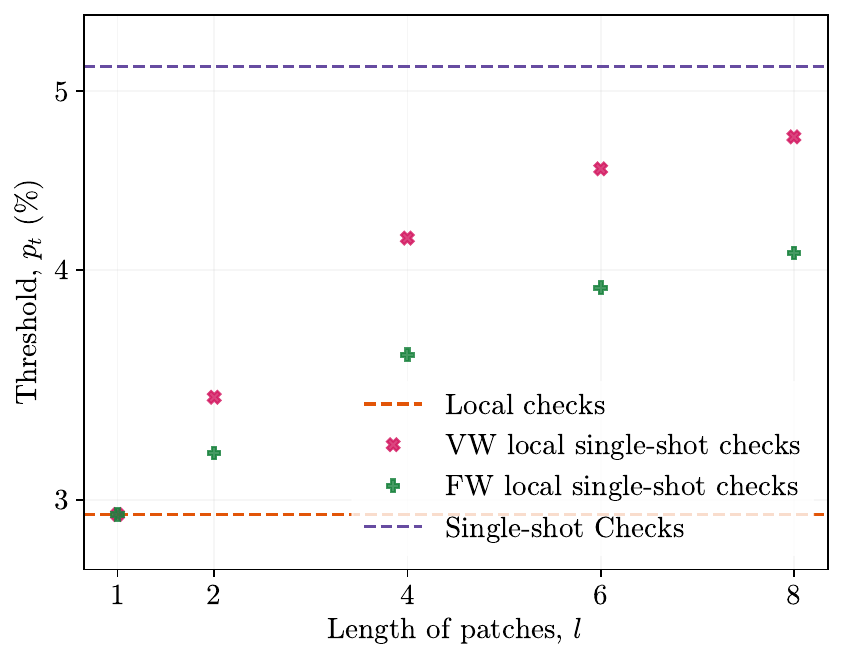}
    \caption{The threshold values of local single-shot checks defined on partitions with different $l$ under a phenomenological noise model. }
    \label{fig:thres_vs_l}
\end{figure}

The detailed relation between threshold and $l$ for both variable-width checks and fixed-width checks are shown in Fig.~\ref{fig:thres_vs_l}. 
When $l=1$, both checks reduce to local checks. 
When $l=L$, the variable-width checks become single-shot checks. 
As a result, the threshold of variable checks converges to that of single-shot checks. 
For the fixed-width checks, when $l=L$, the length of each partition spans one direction of the torus, but the width of the partition is still restricted to $1$. 
This restriction leads to the lower threshold limit for fixed-width checks. 
The threshold for variable-width checks are higher than that of fixed-width checks of the same $l$.
Additionally, as shown in Fig.~\ref{fig:phenom} (center and right), the logical error rates near threshold region of local single-shot checks defined with higher $l$ are lower. 
However, we expect that at low physical error rate $p$, the logical error rates of checks with lower $l$ outperform the ones with higher $l$ as the slopes of the curves are larger for checks with higher $l$.

\begin{figure}[htbp!]
    \centering
    \includegraphics[width=0.9\linewidth]{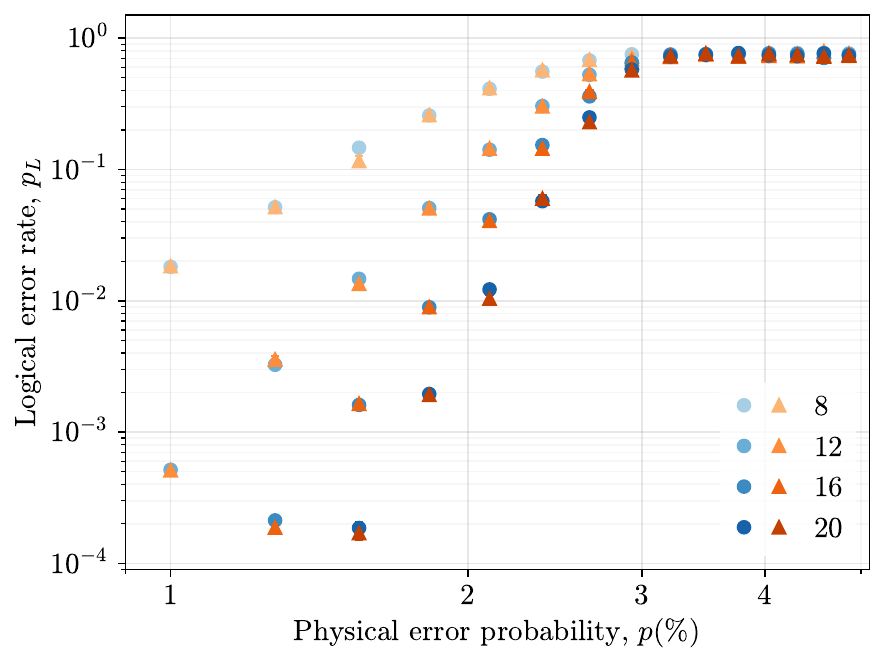}
    \caption{Sliding-window decoding (orange triangles) and whole-syndrome-history decoding (blue circles) of the variable-width checks using the offset measurement scheme with $20d$ rounds of syndrome extraction. The window size $W=d/2+1$. The legend indicates the code distance.}
    \label{fig:sliding_window_phenom}
\end{figure}

We also investigate sliding-window decoding with the local single-shot checks. 
We simulate $20d$ rounds of syndrome extraction for the variable-width checks of $l=2$ and use a window size $W=d/2+1$. 
As shown in Fig.~\ref{fig:sliding_window_phenom}, decoding using a sliding-window decoder with a window size $W=d/2+1$ has almost exactly the same logical error rate as the case where we decode with all the syndrome from the $20d$ rounds of extraction. 
This result confirms that our optimal decoding window that can achieve the best accuracy can be reduced to at most $O(d/2)$. 
We present a comparison of using offset and aligned measurement scheme under the sliding window decoding in App.~\ref{App:offse_vs_daligned}.

\subsection{Circuit-level noise model}{\label{sec:circuit_result}}
We perform circuit-level simulations using Stim \cite{gidney2021stim} and Quits \cite{kang2025quits}. We only focus on the local single-shot checks results obtained with the fixed-width checks. 
Based on the simulation results with the phenomenological noise model, we adopt the offset measurement scheme for all fixed-width checks simulations. 
Every circuit corresponds to $R$ rounds of syndrome extraction, preceded by one noisy round of syndrome extraction used for state preparation, and terminated by a transversal measurement of all data qubits.

\begin{figure*}[htbp!]
    \centering
    \includegraphics[width=0.99\linewidth]{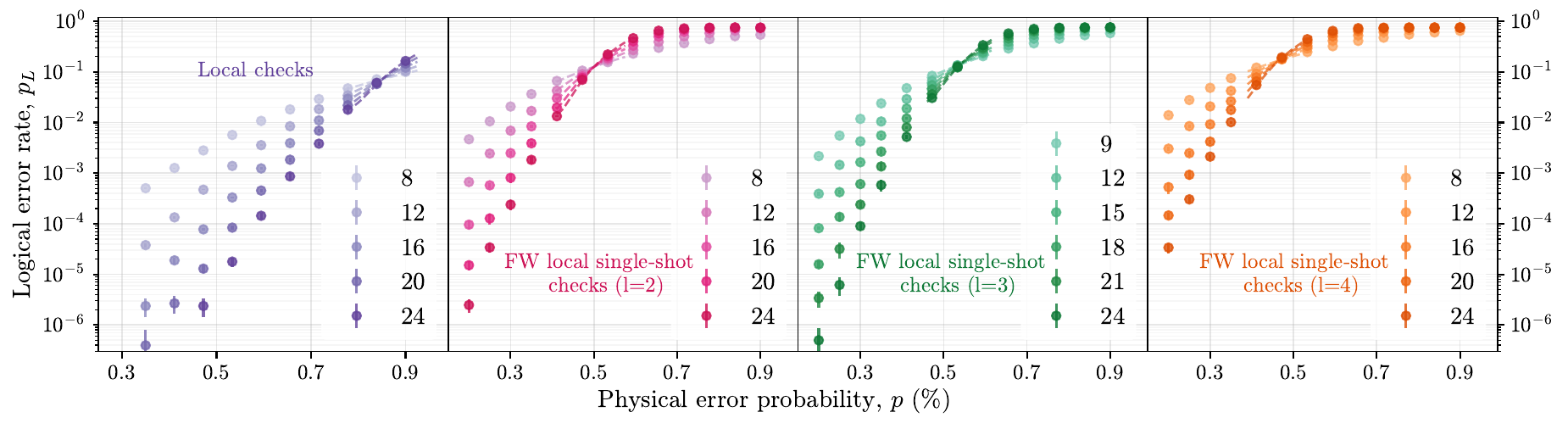}
    \caption{Circuit-level simulation of local checks and fixed-width checks defined on partitions of $l=2,3,4$ using $R=d$ rounds of syndrome extraction. The fixed-width checks are decoded using a decoding constructed from space-edge-first decomposition. The threshold values corresponding to the local and fixed-width checks defined on partitions of $l=2,3,4$ are $0.85\%, 0.50\%, 0.54\%$ and $0.46\%$, respectively. The legend indicates the code distance.}
    \label{fig:circuit_results}
\end{figure*}

A comparison of the space-edge-first decoder and the time-edge-first decoder is shown in App.~\ref{App:complete_vs_reduced}.
The time-edge-first decoder reduces the time distance to that of standard local checks, preventing reduction in the number of syndrome extraction rounds needed for each decoding window. 
Therefore, although for a $O(d)$ window, the time-edge-first decoder achieves a higher threshold and lower logical error rate, we focus on space-edge-first decoder in this section. This decoder preserves the time distance and is thus compatible with reduced-window decoding.

We plot the decoding performance of local checks and fixed-width checks of different $l$ with $R=d$ rounds of syndrome extraction in Fig.~\ref{fig:circuit_results}. 
In this case, the local checks outperform the fixed-width checks in both logical error rate and threshold. 
Nevertheless, the fixed-width checks exhibit a threshold for all cases. 
This stands in contrast with the single-shot checks, for which no threshold exists because their check weight grows unbounded with the code distance, leading to unbounded measurement error rates \cite{lin2024single}.

As $l$ increases from $2$ to $3$, the logical error rate decreases and the threshold increases slightly. 
This behavior can be attributed to the fact that the density of time edges is lower for $l=3$ compared to $l=2$, which means that the data qubit errors in $l=3$ have a lower probability of becoming a long-lived error chain spanning the boundary of the decoding window. Additionally, the weights of the checks remain reasonably low to not offset the improvement in decoding performance. 
However, increasing to $l=4$, we observe that the threshold decreases slightly and the logical error rate becomes much higher. 
Apart from the fact that more high-weight checks are involved, the internal error structures that emerge for $l=4$ correlate detectors in a way that cannot be decomposed into one space edge with a measurement error on a fixed-width check. These structures lead to a greater loss of correlation information during decomposition and thus degrade performance. 
The detailed discussion of the space-edge-first decomposition can be found in App.~\ref{App:decomp}.

\begin{figure*}[htbp!]
    \centering
    \includegraphics[width=0.99\linewidth]{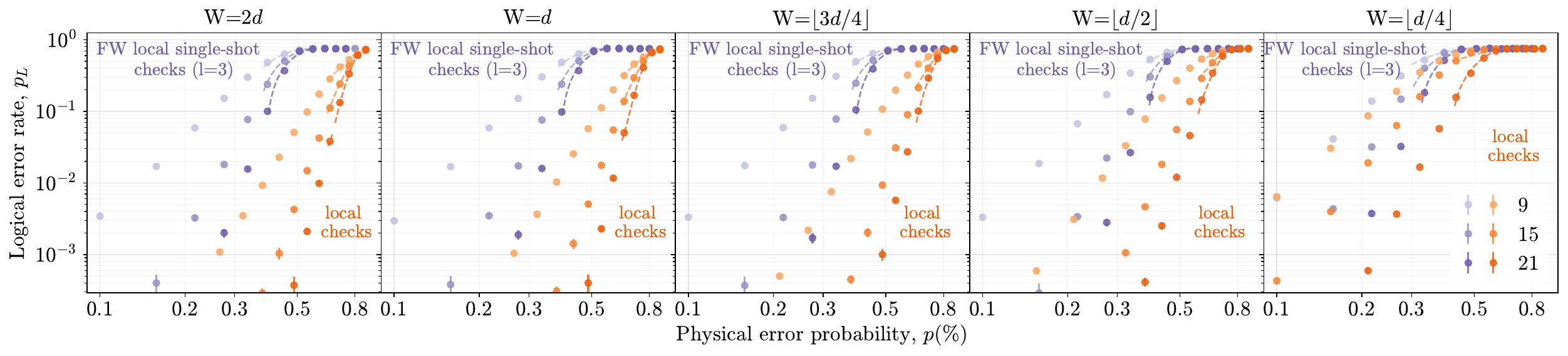}
    \caption{Circuit-level simulation of  local checks and fixed-width (FW) local single-shot checks with $l=3$ using sliding-window decoding. The total length of syndrome extraction is $20d$. We vary the window size $W$ from $2d$ to $\lfloor d/4\rfloor$. The threshold values for the local checks for window sizes $2d, d, \lfloor3d/4\rfloor, \lfloor d/2\rfloor$ and $\lfloor d/4\rfloor$ are $0.82\%, 0.82\%, 0.80\%, 0.74\% $ and $0.62\%$, respectively. The threshold values for the fixed-width local single-shot checks on a patch of size $l=3$ at the same window sizes are $0.56\%$, $0.58\%$, $0.56\%$, $0.56\%$ and $0.46\%$, respectively. The legend indicates the code distance.}
    \label{fig:threshold_vs_window_size}
\end{figure*}

Based on the results in Fig.~\ref{fig:circuit_results}, we choose checks of $l=3$ for the sliding-window simulations. 
We run circuits with $R=20d$ syndrome extraction rounds. 
We present sliding-window decoding results for both local checks and the fixed-width checks with $l=3$ using multiple different window size $W$ in Fig.~\ref{fig:threshold_vs_window_size}. 
To study the role of time distance $d_t$, we also include cases where $W$ is smaller than the lower bound given in Eq.~\ref{eq:optimal}. 

The local checks exhibit a higher threshold among all the $W$ we simulate. 
Specifically, when $W=2d$ and $W=d$, the threshold of local checks remains at $\sim0.82\%$. 
The threshold starts to drop slowly when window size reduces below $d$. 
When $W=\lfloor d/4\rfloor$, the threshold reduces to $\sim0.62\%$. 
At the same time, when $W$ decreases from $2d$ to $d$, the logical error rate and slopes remain nearly unchanged, which confirms that $W=O(d)$ is sufficient for fault-tolerant sliding-window decoding. 
As $W$ drops below $d$, the logical error rates begin to increase and the slope begins to reduce. 
This performance degradation numerically confirms that $d_{eff}=O(d)$ is optimal for a decoding window. 

For the fixed-width checks, when $W\ge d$, the logical error rate is higher than the local checks. 
This results from many factors, including an increase due to mapping the measurement errors into data qubit errors, which also increases the logical error rate in the phenomenological noise model, as well as the effects of high-weight checks and non-optimal circuits. 
Notably, the non-optimal decoding techniques to handle the internal errors significantly contribute to the overall logical error rate. 
See App.~\ref{App:complete_vs_reduced} for a comparison of different hyperedge decompositions. 
However, the decoding behavior of fixed-width checks is more robust against the decoding window size. 
The logical error rate remains nearly unchanged as $W$ decreases from $2d$ to $\lfloor 3d/4\rfloor$, and only increases modestly for $W \le \lfloor d/2\rfloor$.
The slope also remains nearly constant for $W \ge \lfloor d/2\rfloor$, and only begins to decrease when $W = \lfloor d/4\rfloor$.

This means that we can use less syndrome extraction rounds within a decoding window for the fixed-width checks to achieve its optimal decoding performance. 
As $W$ decreases, the curves of local checks and fixed-width checks start to overlap. 
When $W=\lfloor d/4\rfloor$, the slopes of local check curves are lower than that of the fixed-width checks, so that the fixed-width checks outperform the local checks at low physical error rate. 

\begin{figure}[htbp!]
    \centering
    \includegraphics[width=0.99\linewidth]{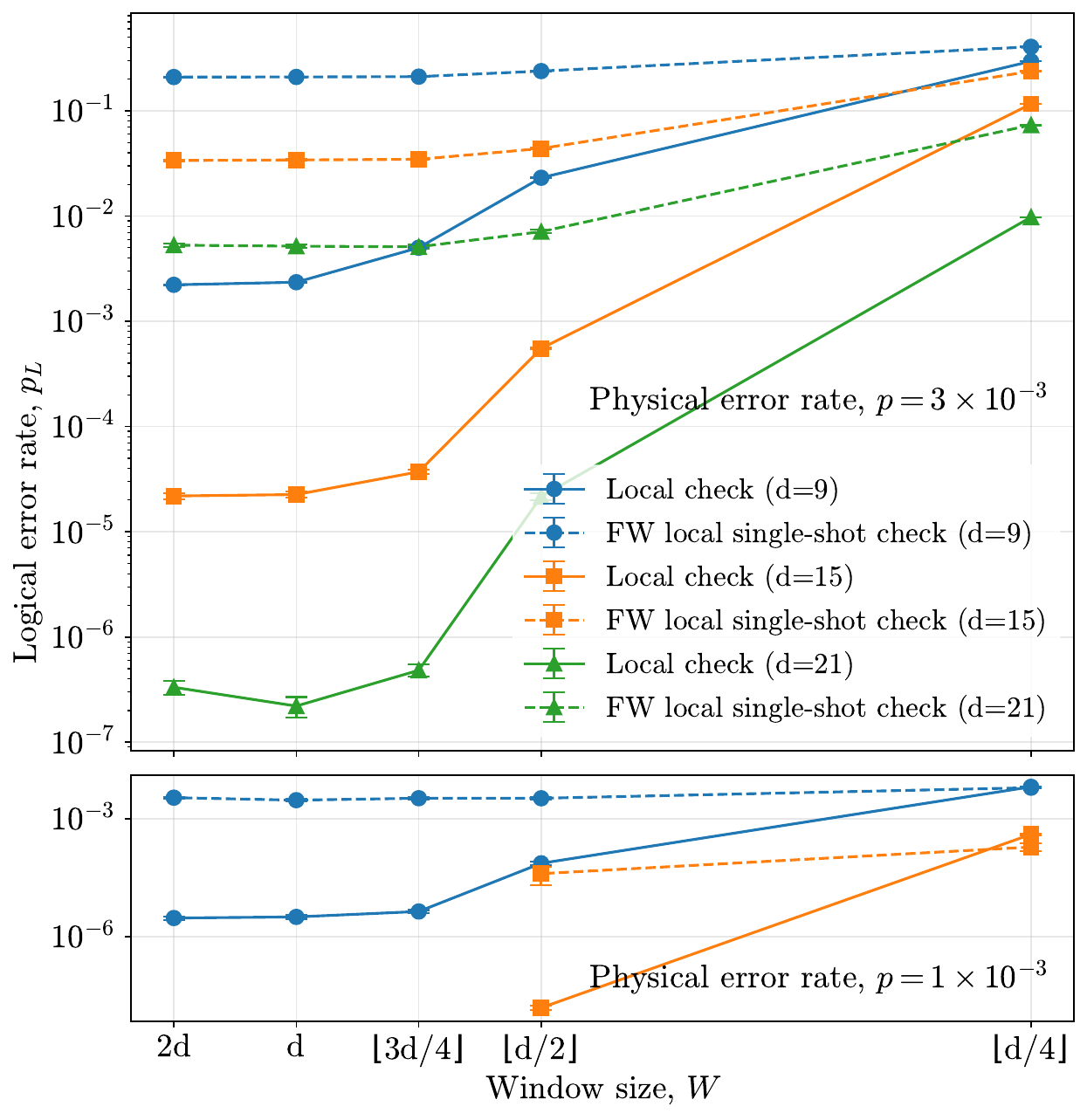}
    \caption{Logical error rate as a function of window size for local checks and the fixed-width checks of $l=3$. Dashed and solid lines represent fixed-width local single-shot checks and local checks, respectively. The legend indicates the code distance.}
    \label{fig:ler_ecd_vs_ws}
\end{figure}

Finally, in Fig.~\ref{fig:ler_ecd_vs_ws}, we plot the logical error rate as a function of $W$ at two representative physical error rates ($p=1\times 10^{-3}$ and $3\times 10^{-3}$) for both local checks and local single-shot checks.
At $p = 3\times 10^{-3}$, the logical error rate for local checks remains nearly constant for $W \ge d$, and increases sharply for $W < d$. 
In contrast, for the local single-shot checks, the logical error rate is relatively robust against the window size, with a slight increases observed at $W\le \lfloor d/2\rfloor$.
As we can see in the lower subplot, at low $p=10^{-3}$, the curves of local checks and local single-shot checks start to cross at $W=\lfloor d/4\rfloor$. 
This indicates that our local single-shot checks has an advantage in decoding performance under low $W$ compared to the local checks due to the larger time distance within a given decoding window of $W$.
We also expect crossing of curves to happen for lower error rate for other $W<d$. 
This prediction is based on the fact that, at low $p$, the logical errors are dominated by unrecoverable error chains of the minimum weight, which is upper bounded by the time distance. 

We also present the results of sliding-window decoding with different $l$ in Fig.~\ref{fig:Patch_size_comparision} and a detailed discussion in App.~\ref{App:circuit_l_compare}.
We emphasize that neither the circuit nor the decoder we use in this section is optimal. 
Furthermore, there are many possible choices for constructing local single-shot checks given different partition of the surface and various choices of local single-shot checks within a patch. 
We expect that improved circuit design and decoding strategies would further enhance performance.

\section{Conclusion}\label{sec: conclusion}
We introduced local single-shot checks for the toric code by partitioning the lattice into small patches.
By choosing a dynamic measurement scheme, we reduce the number of rounds required within an optimal sliding-window decoder can be reduced by a factor that depends on the patch size.
We numerically confirm this property with circuit-level simulation using a sliding-window decoder. 

The idea of local single-shot checks proposed in this paper can have various constructions based on different partitions of the torus. 
In this work, we present two different choices of local single-shot checks, the variable-width checks and the fixed-width checks based on a square or rectangular partition of the torus. 
Even within the patches of a given partition, the choice of local single-shot checks can be different. 
For example, we can choose different set of data qubits to map the measurement errors of local single-shot checks onto. 
These choices of local single-shot constructions can have different advantages under various noise model. 
As shown in this paper, the fixed-width checks can be implemented by syndrome extraction with bare syndrome qubits that preserves the effective circuit distance. 
The variable-width checks show advantages in the phenomenological noise model with higher thresholds and lower logical error rate given the length of the patches. 
We believe optimizing the construction of local single-shot checks according to noise models may further boost the code performance.

The time distance, which is the focus of this paper, might inspire various designs that balance the logical accuracy and the syndrome extraction rounds. 
For example, if we mix the measurement of local checks and the local single-shot checks, we might be able to achieve lower logical error rate with reduced measurement rounds.

Our simulation also reveals that internal errors can be destructive for single-shot-based check design aiming at reducing measurement overhead as they can introduce diverse hyperedge structures and potentially reduce effective code distance during decoding. 
For the circuit we design empirically, we provide two MWPM-based decoders using time-edge-first hyperedge decomposition and space-edge-first hyperedge decomposition respectively. 
During the hyperedge decomposition, we naively increase the weight of the edges in the decoding graph by combining the probability of error events. 
Although our decoders already achieve good performance, further improvements are possible by optimizing syndrome extraction circuits to avoid harmful internal error sources and optimizing the decoder itself, e.g. by optimizing the weight distribution and edge structures of the decoding graph \cite{higgott2023improved} or by employing hyperedge-aware decoders \cite{roffe_decoding_2020,beni2025tesseractdecoder}.

Finally, while our study focused on the toric code, the notion of local single-shot checks and time-distance-aware decoding applies more broadly.
We expect our construction can be generalized to other codes \cite{shor1995scheme, bacon2006operator,li20192d} where the time distance can be clearly defined. 
We expect that interesting results will occur studying other code families.
Exploring these directions may yield new pathways for reducing measurement overhead while maintaining high fault-tolerance performance.

\section*{Acknowledgment}
The authors thank Shilin Huang, Balint Pato and Julie Campos for helpful discussion.
This work was supported by the National Science Foundation (NSF) Quantum Leap Challenge Institute for Robust Quantum Simulation (QLCI, Grant No. OMA-2120757), and the Office of the Director of National Intelligence (ODNI) Intelligence Advanced Research Projects Activity (IARPA) under the Entangled Logical Qubits program through Cooperative Agreement No. W911NF-23-2-021.

\bibliography{main}

\newpage

\onecolumngrid
\appendix

\section*{Appendix}

\subsection{Hyperedge decomposition for circuit-level decoding}\label{App:decomp}
Here we introduce the time-edge-first and space-edge-first decomposition used to construct decoding graphs for the circuit-level noise model in detail.
Note that whenever a hyperedge is decomposed into multiple edges, we assign each resulting edge the same weight as the original hyperedge.
If the edge already exists in the graph, we update the weight of these edges based on the joint probability of the edge being turned on by different error sources. 
There are three classes of errors in the circuit, flipping $2$, $4$ and $6$ detectors respectively. 
These error mechanisms give rise to degree-$2$ edges, degree-$4$ hyperedges and degree-$6$ hyperedges, as discussed in Sec.~\ref{sec:circuit-decoder}. 

\subsubsection{Time-edge-first hyperedge decomposition}\label{app:time-edge}

In the time-edge-first hyperedge decomposition, we directly add all edges into the decoding graph. 
Below, we outline how degree-$4$ and degree-$6$ hyperedges are handled. 

\begin{enumerate}
    \item {\textbf{Degree-$4$ hyperedges}:}
    
    There are two types of hyperedges in this class.
    \begin{itemize}
        \item \emph{Hyperedges containing two pairs of detectors that share spatial coordinates. }
        
        An example is given in Fig.~\ref{fig:decomposition_meas}(a).
        These type of hyperedges are typically arise from measurement errors on top or bottom-boundary checks. 
        We decompose them into two time edges as is shown in Fig.~\ref{fig:decomposition_meas}(c).
        
        \item \emph{Hyperedges containing only one pair of detectors that share spatial coordinates but are located on adjacent time layers.}
        
        In such cases, the remaining two detectors can be on the same or different time layers.
        In both cases, we add a time edge between the pair of detectors sharing the same spatial coordinates and add an edge connecting the remaining two detectors.
        Examples can be found in Fig.~\ref{fig:time-edge-first}(a) and (b). 
        Fig.~\ref{fig:time-edge-first}(a) shows the case where the remaining detectors are on different time layers, so we add a space-time edge between them. Fig.~\ref{fig:time-edge-first}(b) shows the case where the remaining detectors are in the same time layer, so we add a space edge between them.
        
    \end{itemize}

    \begin{figure*}[htbp!]
    \centering
    \includegraphics[width=0.5\linewidth]{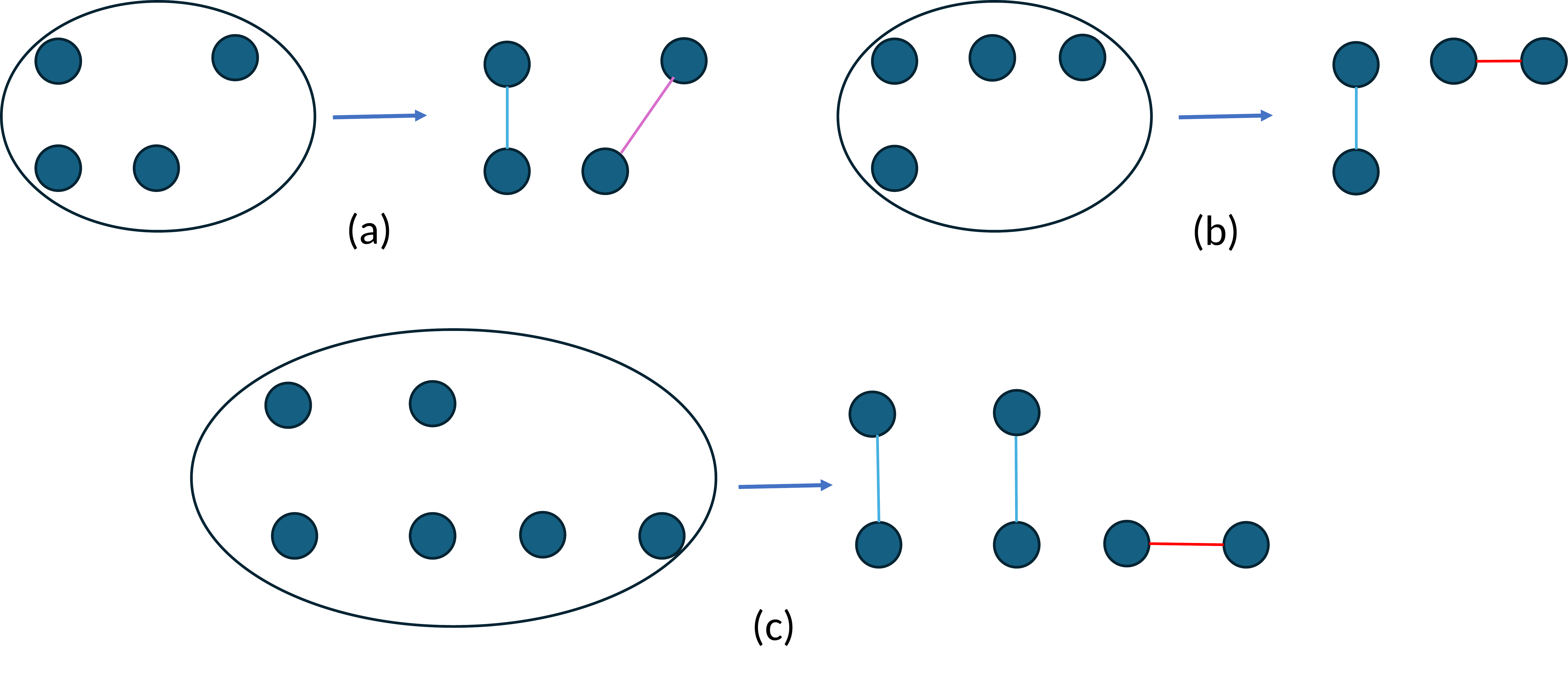}
    \caption{Illustration of time-edge first decomposition. All these hyperedges (represented by black circles) are caused by internal errors. The rows in each subplot from top to down representing time layer $t+1$ and $t$. The relative positions within a row represent the spatial coordinate. Blue edges are time edges. Red edges are space edges. Purple edges are space-time edges. (a) A degree-$4$ hyperedge is decomposed into a time edge and a space-time edge. (b) A degree-$4$ hyperedge is decomposed into a time edge and a space edge. (c) A degree-$6$ hyperedge is decomposed into two time edges and a space edge.}
    \label{fig:time-edge-first}
\end{figure*}

    \item {\textbf{Degree-$6$ hyperedges}:}

    These edges typically involve two pairs of detectors sharing spatial coordinates, along with two additional detectors. 
    We add two time edges between the two pairs and an additional edge to connect the remaining detectors. 
    An example is shown in Fig.~\ref{fig:time-edge-first}(c) where we add two time edges and one space edge to decompose the hyperedge.
\end{enumerate}

The time-edge-first decomposition does not restrict the allowed time edges, so the time distance of the decoding window remains equal to the number of syndrome extraction rounds in the window.
Therefore, we do not expect any measurement reduction using decoding graphs based on this decomposition.

\subsubsection{Space-edge-first hyperedge decomposition}\label{app:space-edge}

In the space-edge-first decomposition, we aim to preserve the same time distance across different time layers as in the phenomenological noise model.
To achieve this, we impose the constraint that the allowed space-time edges and time edges can only start from the allowed starting points $(i,j,t)$ and end at the allowed ending position $(i,j',t+1)$ of the same spatial row in the next layer $t+1$, where $j$ and $j'$ can be different. 
In the following, we detail how different types of edges and hyperedges are handled. To highlight the role of internal errors, we distinguish hyperedges that occur for all simulated $l$ from those that appear only when $l=4$ in the discussion below.

\begin{enumerate}
    \item {\textbf{Edges}:}
    \begin{itemize}
        \item \emph{No auxiliary detectors needed.}

        All edges from $l=3,4$ and most edges of $l=2$ belong to this class. 
        These edges are added directly to the decoding graph.

        \item \emph{Auxiliary detectors needed.}

        For certain $l=2$ cases, an edge may connect detectors from allowed starting points $(i,j,t)$ to disallowed ending points $(i',j',t+1)$. 
        Adding these edges directly will reduce the time distance between layers. 
        These edges are typically introduced by internal errors. 
        For example, an internal error can happen after measuring the top boundary checks $\tilde{T}^{u+1,v}$ and the full-patch checks $\tilde{F}^{u+1,v}$ but before the full-patch checks $\tilde{F}^{u,v}$ on the qubit that is shared by the two patches labeled by $\{u,v\}$ and $\{u+1,v\}$. 
        To handle these edges, we insert an auxiliary detector at the allowed ending point $(i,j,t+1)$ of the time edges. 
        We can then decompose the original edge into a time edge connecting $(i,j,t)$ and  $(i,j,t+1)$, and a space edge connecting  $(i,j,t+1)$ and  $(i',j',t+1)$, as  $(i,j,t+1)$ and $(i',j',t+1)$ are always local for the circuit we study. 
        This corresponds to interpreting the internal error as a measurement error on $\tilde{F}^{u,v}$ at time $t$ followed by a data qubit error at time $t+1$.  
        An example is shown in Fig.~\ref{fig:decomposition_meas}(d) and (e).
    \end{itemize}

    \item {\textbf{Degree-4 hyperedges}:}
    
    \begin{itemize}
        \item \emph{Two pairs of detectors share spatial coordinates.}

        An example is given in Fig.~\ref{fig:decomposition_meas}(a). 
        These hyperedges are typically caused by measurement errors. 
        We decompose them into two space edges as in the phenomenological noise model (see Fig.~\ref{fig:decomposition_meas}(b)). 
        With this decomposition, we map them onto data qubit errors happening at time $t$ and $t+1$. 

        \item \emph{One pair of detectors share the same spatial position.} \begin{itemize}
            \item \textit{The pair of detectors lies on allowed time edge position and the remaining two detectors are on the same time layer. }

            We introduce a time edge between the spatially aligned detectors and a space edge to connect the remaining detectors. 
            An example of this decomposition can be found in Fig.~\ref{fig:space-edge-first}(a). 
            These errors can also be caused by internal errors. 
            For example, an error can occur at the right boundary qubit of the top boundary check $\tilde{T}^{u,v}$ right before the last full-patch check $\tilde{F}^{u,v}$ supported on the same qubit is measured. 
            Therefore, the error will only flip $\tilde{F}^{u,v}$ in the current syndrome extraction round. 
            However, the next round of syndrome extraction will capture the complete syndrome of this error that involves two top boundary checks $\tilde{T}^{u,v}$  and $\tilde{T}^{u,v+1}$  and two full-patch checks $\tilde{F}^{u,v}$ and  $\tilde{F}^{u,v+1}$. 
            We therefore interpret such error as a measurement error on $\tilde{F}^{u,v}$ with the actual data qubit error.

\begin{figure*}[t]
    \centering
    \includegraphics[width=0.5\linewidth]{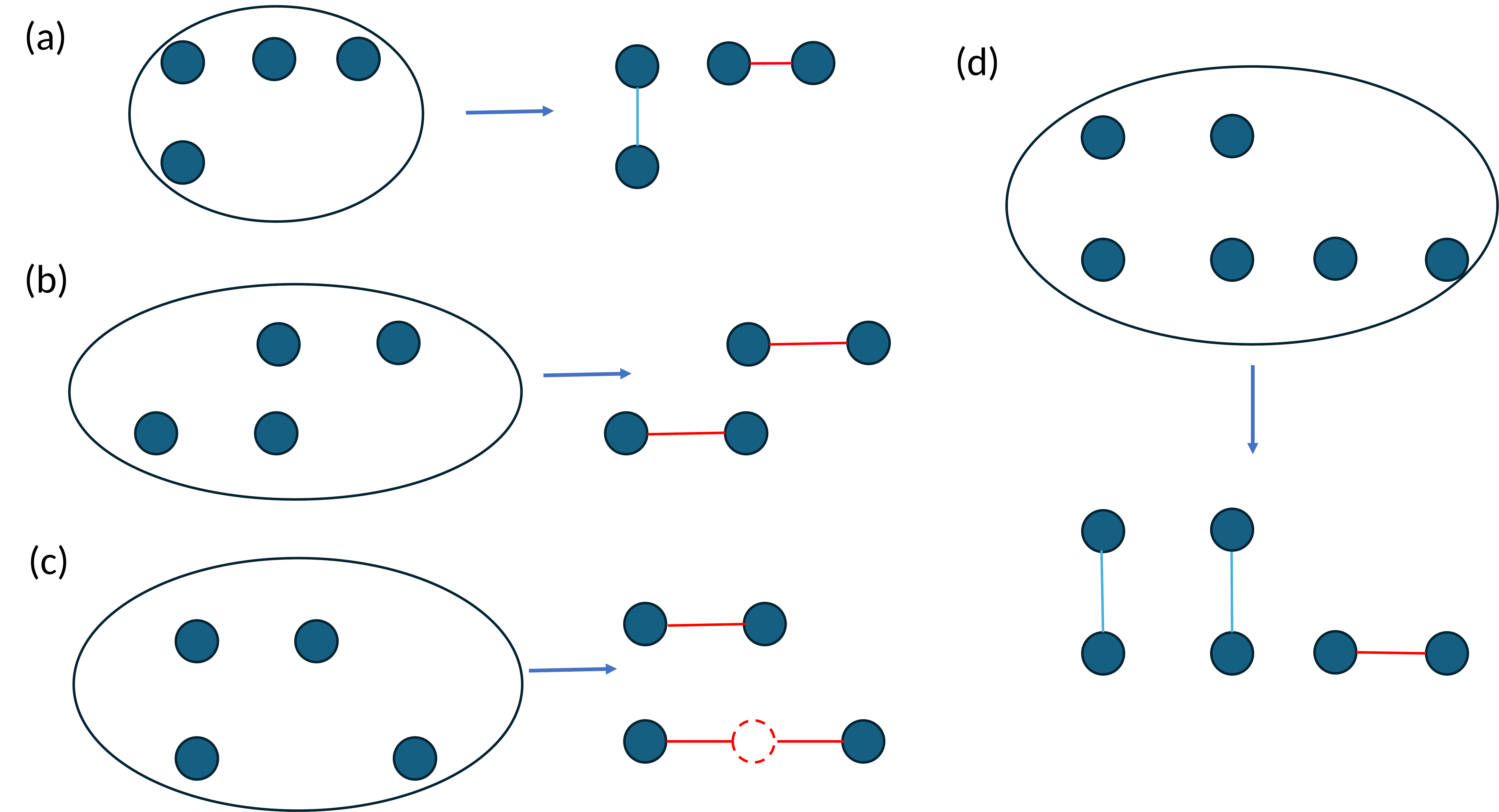}
    \caption{Illustration of space-edge-first decomposition that applies to all $l$. All these hyperedges (represented by black circles) can be caused by internal errors. The rows in each subplot from top to down representing time layer $t+1$ and $t$. The relative positions within a row represent the spatial coordinate. Blue edges are time edges. Red edges are space edges. Dashed dots represent auxiliary detectors we add. Blue dots represent detectors that are in support of the hyperedge. (a) A degree-$4$ hyperedge is decomposed into a time edge and a space edge. (b) A degree-$4$ hyperedge is decomposed into two space edges that are not aligned in space. (c) A degree-$4$ hyperedge is decomposed into three hyperedges with one auxiliary detector inserted. (d) A degree-$6$ hyperedge is decomposed into two time edges and one space edge.}
    \label{fig:space-edge-first}
\end{figure*}

        \item \emph{The remaining two detectors lie on different time layers.}

        \begin{itemize}
            \item \emph{Each time layer contains two detectors that are local. }
            
            In this case, we add space edges between detectors on the same time layers. 
            An example is given in Fig.~\ref{fig:space-edge-first}(b). 
            Such errors are typically internal errors occurring on the left boundary of the patch labeled by $(u,v)$ right after the first check that has support on this qubit, the top boundary check  $\tilde{T}^{u,v}$, is measured. 
            The syndrome is equivalent to that of a measurement error on $\tilde{T}^{u,v}$ and an error on the same data qubit that occurs before the syndrome extraction round.
            
            \item \emph{Each time layer contains two detectors that can be nonlocal.}

            In this case, we cannot directly decompose the hyperedge into two local space edges. 
            Instead, we insert an auxiliary detector by projecting one of the detectors $(i,j,t+1)$ that do not share spatial coordinate with any other detectors supporting this hyperedge to $(i,j,t)$ on the hyperedge's other time layer. 
            In this way, we can always connect the auxiliary detector with two other detectors on the time layer $t$ space edges and used another space edge to connect the remaining detectors on the time layer $t+1$ of the hyperedge. 
            An example is shown in Fig.~\ref{fig:space-edge-first}(c).
            This decomposition is essentially interpreting the internal error on the top boundary of the patch labeled by $(u,v)$ happening right after the measuring the top boundary check $\tilde{T}^{u,v}$ supported by the same qubit into a measurement error on that boundary check and a data qubit error.
        \end{itemize}
        \end{itemize}

    \end{itemize}
    
    \item {\textbf{Degree-6 hyperedges}:}
    
    Here we only consider the hyperedges that have two pairs of detectors sharing the same spatial position and two additional detectors that are spatially local to each other. 
    These are the only degree-$6$ hyperedges for $l=2,3$. 
    In this case, since the two pairs of detectors always lie in the allowed positions for time edges, we can decompose them into time edges and connect the remaining detectors with a space edge as all of the remaining detectors in this case lie locally in the same time layer.
    An example is shown in Fig.~\ref{fig:space-edge-first}(d). 
    These hyperedges can be caused by internal errors that occur on the left or right boundary of each patch in support of either top or bottom boundary checks before we start measuring the full-patch checks. 
    This decomposition is performed in the same manner as in the time-edge-first decoder.

\begin{figure*}[t]
    \centering
    \includegraphics[width=0.5\linewidth]{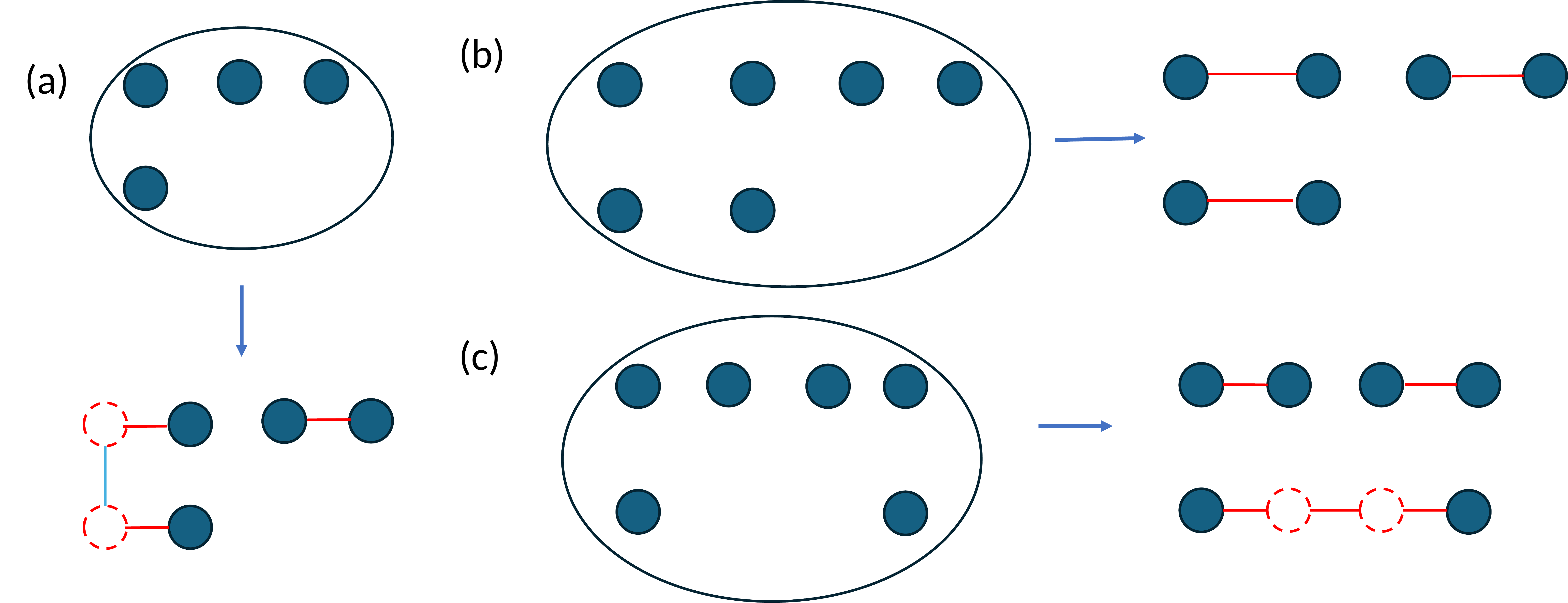}
    \caption{Illustration of space-edge-first decomposition for cases only exist when $l=4$. The black circles represent hyperedges. Blue dots are detectors in support of the hyperedges. Red dashed dots are auxiliary detectors we insert to decompose hyperedges. The rows in each subplot from top to down representing time layer $t+1$ and $t$. The relative positions within a row represent the spatial coordinate. Blue edges represent time edges. Red edges represent space edges. (a) A degree-$4$ hyperedge is decomposed into three space edges and one time edge with two inserted auxiliary detectors. (b) A degree-$6$ hyperedge is decomposed into three space edges at both time layers. (c) a degree-$6$ hyperedge is decomposed into five space edges with two auxiliary detectors inserted.}
    \label{fig:l4decomp}
\end{figure*}

    \item {Hyperedges only for $l=4$:}

    \begin{itemize}
        \item \textbf{Degree-4 hyperedges.}

        For $l=4$, there can be cases where the pair of detectors $(i,j,t)$ and $(i,j,t+1)$ that share the spatial coordinate do not locate at the allowed time edge positions while the other two detectors are local on the same time layer. 
        We connect the other two detectors by a space edge. For the pair of detectors $(i,j,t)$ and $(i,j,t+1)$, we insert a pair of auxiliary detectors that locate at the nearest allowed time edge position $(i-1,j,t)$ and $(i-1,j,t+1)$.
        We then use space edges to connect $(i-1,j,t)$ and $(i,j,t)$. $(i-1,j,t+1)$ and $(i,j,t+1)$. 
        Additionally, we introduce a time edge to connect $(i-1,j,t)$ and $(i-1,j,t+1)$. 
        An example is shown in Fig.~\ref{fig:l4decomp}(a).
        These hyperedges are caused by internal errors that happen on the bottom boundary qubit of the patch labeled by $(u,v)$ before the final two checks $\tilde{B}_{2}^{u,v}$ and $\tilde{F}^{u,v}$ that have support on them are measured. The error syndrome is equivalent to that of a measurement error on $\tilde{F}^{u,v}$ and $\tilde{B}_{2}^{u,v}$ together with a data qubit error on the bottom boundary qubit of the pathch $(u,v)$.

        \item \textbf{Degree-6 hyperedges.}

        For $l=4$, there exist cases where at least one of the two pair of detectors that share the spatial coordinate do not lie at allowed time edge positions. 
        We can further divide these edges into the following two classes.
        \begin{itemize}
            \item \emph{The two pairs of detectors are spatially local, while the remaining two detectors are spatially local and lie on the same time layer.}
            
            In this case, we decompose the hyperedges into three space edges, two connecting the two detectors of the same time layers inside the two pairs and one connecting the remaining two detectors. 
            An example is given in Fig.~\ref{fig:l4decomp}(b). 
            Such hyperedges can occur due to internal errors. 
            For example, an internal error can occurs at the left boundary qubit of the bottom boundary check $\tilde{B}^{u,v}_{l-1}$ right before the measuring the bottom boundary checks $\tilde{B}^{u,v}_{2}$ and $\tilde{B}^{u,v-1}_{2}$ and full-patch check $\tilde{F}^{u,v}$ and $\tilde{F}^{u,v-1}$. 
            The error syndrome can be interpreted as measurement errors on these four checks, as well as a data qubit error.

            \item \emph{The two pairs of detectors are not spatially local. }
            
            In such cases, there are always four detectors on the $t+1$ layer and we can connect them by space edges.
            For the remaining two detectors on the $t$ layer, we introduce two auxiliary detectors so that we can connect them all with space edges. 
            An example can be found in Fig.~\ref{fig:l4decomp}(c).
            Such hyperedges are typically caused by internal errors. 
            For example, an internal error can happen on the top boundary qubit of the patch labeled by $(u,v)$ after the measurement of the top boundary check $\tilde{T}^{u,v}$ and the bottom boundary check $\tilde{B}^{u-1,v}_{l-1}$.
            The syndrome of this error can be interpreted as measurement errors on these two checks and a data qubit error on the top boundary qubit of patch $(u,v)$.

        \end{itemize}
    \end{itemize}
\end{enumerate}

With the restrictions imposed on allowed time edges, the space-edge-first decomposition maintains the same time distance as the phenomenological noise model.
As a result, we can reduce the syndrome extraction rounds for each decoding window. 
Note that whenever we perform hyperedge decomposition, especially when we introduce auxiliary detectors, the decoding graph loses the correlation information of detector flips. 
Therefore, we expect an increase in logical error rate when using this decomposition and a decrease of effective code distance. 
Here, the effective code distance $d_{eff}$ is defined as $d_{eff}=2t_E+1$ where $t_E$ is the maximum weight of errors that are guaranteed to be correctable using a given decoder under a chosen circuit with a specific noise model.

\subsection{A comparison between offset and aligned measurement schemes}\label{App:offse_vs_daligned}
We present numerical results for a comparison between the offset measurement scheme and the aligned measurement scheme under the phenomenological noise model.

\begin{figure}[htbp!]
    \centering
    \includegraphics[width=0.475\linewidth]{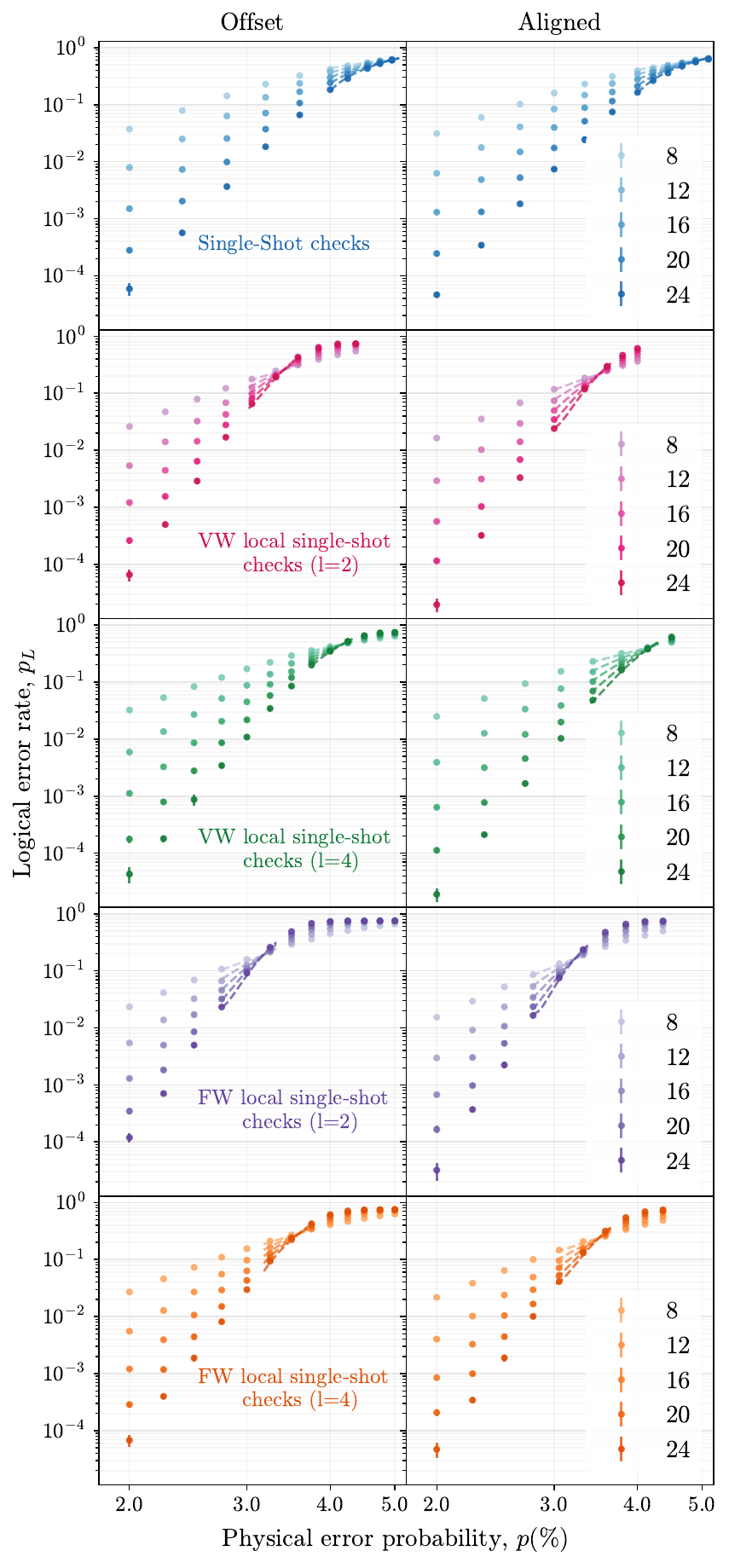}
    \caption{Comparison of logical error rates between the offset and aligned measurement schemes for single-shot, variable-width (VW) local single-shot checks, and fixed-width (FW) local single-shot checks at fixed $l$. The legend indicates the code distance.}
    \label{fig:offset_vs_non_offset}
\end{figure}

Figure~\ref{fig:offset_vs_non_offset} shows the results for single-shot checks and both variable- and fixed-width checks at various $l$, for $R=d$ rounds of syndrome extraction. 
Since $O(d)$ rounds of syndrome is used for decoding, we expect the effect of time distance is negligible.
The offset and aligned measurement schemes yield similar threshold values across all cases, as summarized in Table~\ref{tab:thres}.
This indicates that the threshold enhancement primarily arises from the single-shot check structure, rather than the choice of measurement schedule.
In the near-threshold region, the logical error rates are also comparable, with the aligned scheme performing slightly better for small $l$.
This difference results from time edge distribution in the decoding graph.

\begin{table}[htbp!]
    \centering
    \begin{tabular}{c|
                    >{\centering\arraybackslash}p{3cm}|
                    >{\centering\arraybackslash}p{2.7cm}|
                    >{\centering\arraybackslash}p{2.7cm}|
                    >{\centering\arraybackslash}p{2.7cm}|
                    >{\centering\arraybackslash}p{2.7cm}}
        \toprule
         &  Single-shot check 
         & \multicolumn{2}{c|}{Variable-width local single-shot check}
         & \multicolumn{2}{c}{Fixed-width local single-shot check} \\
        \midrule
         \multicolumn{2}{c|}{}  & $l=2$ & $l=4$ &  $l=2$ & $l=4$ \\
        \midrule
        Offset  & $5.16\%$ &  $3.41\%$ &  $4.16\%$ & $3.18\%$ & $3.59\%$ \\ 
        \midrule
        Aligned & $5.00\%$ &  $3.51\%$ &  $4.19\%$ &  $3.22\%$ & $3.50\%$ \\ 
        \bottomrule
    \end{tabular}
    \caption{Threshold values for offset and aligned measurement schedules for single-shot, variable-width, and fixed-width local single-shot checks at patch sizes \(l = 2\) and \(l = 4\).}
    \label{tab:thres}
\end{table}

\begin{figure}[htbp!]
    \centering
    \includegraphics[width=0.75\linewidth]{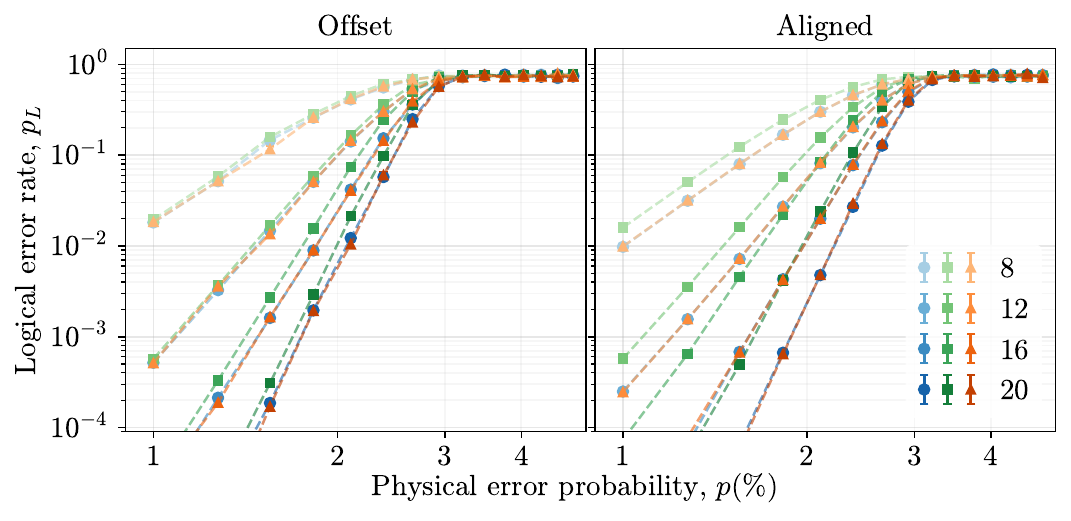}
    \caption{Logical error rates under sliding-window decoding compared to complete-history decoding (blue circles) for the offset (left) and aligned (right) measurement schemes. Window sizes $W=d/2+1$ (orange triangles) and $W=d/4$ (green squares) are shown. The legend indicates the code distance.}
    \label{fig:SW_VS_CH}
\end{figure}

To understand the advantage of an increased time distance of the decoding volume, we further perform sliding window decoding with both measurement schemes using different window sizes. 
Fig.~\ref{fig:SW_VS_CH} shows the results for both measurement schemes, with the complete-history decoding included as a reference.
For both measurement schemes, decoding with complete history (blue circles) gives almost the same logical error rate as using sliding window decoding with a window size $W=d/2+1$ (orange triangles) at near threshold region. 
This indicates that both measurement schemes inherit the robustness of the single-shot check structure, performing effectively in the near-threshold region despite using fewer rounds of syndrome extraction per decoding window. 

However, because the aligned measurement scheme has a smaller time distance than the offset scheme for a given $W<d$, we expect that at low physical error rate, the offset scheme should outperform the aligned scheme. 
To show this difference, numerically, we simulate of the sliding-window decoding using window size $W=d/4$ (green squares) for both cases. 
With this reduced window size, we can see the effect of the difference in time distances more clearly. 
For the offset scheme, sliding-window decoding remains close to complete-history decoding for small distances and only deviates at larger distances.
This gap indicates that we are already below the optimal decoding window size and the effect of the reduced time distance starts to show up.
In contrast, for the aligned scheme, sliding-window decoding exhibits a larger and more persistent gap from complete-history decoding across distances. 
For instance, the $d=20$ curve using $W=d/4$ nearly overlaps with the $d=16$ curve using complete-history decoding.
This confirms that the aligned measurement scheme has a smaller time distance than the offset scheme.

\subsection{A comparison between the time-edge-first and the space-edge-first decomposition}\label{App:complete_vs_reduced}

\begin{figure}[htbp!]
    \centering
    \includegraphics[width=0.44\linewidth]{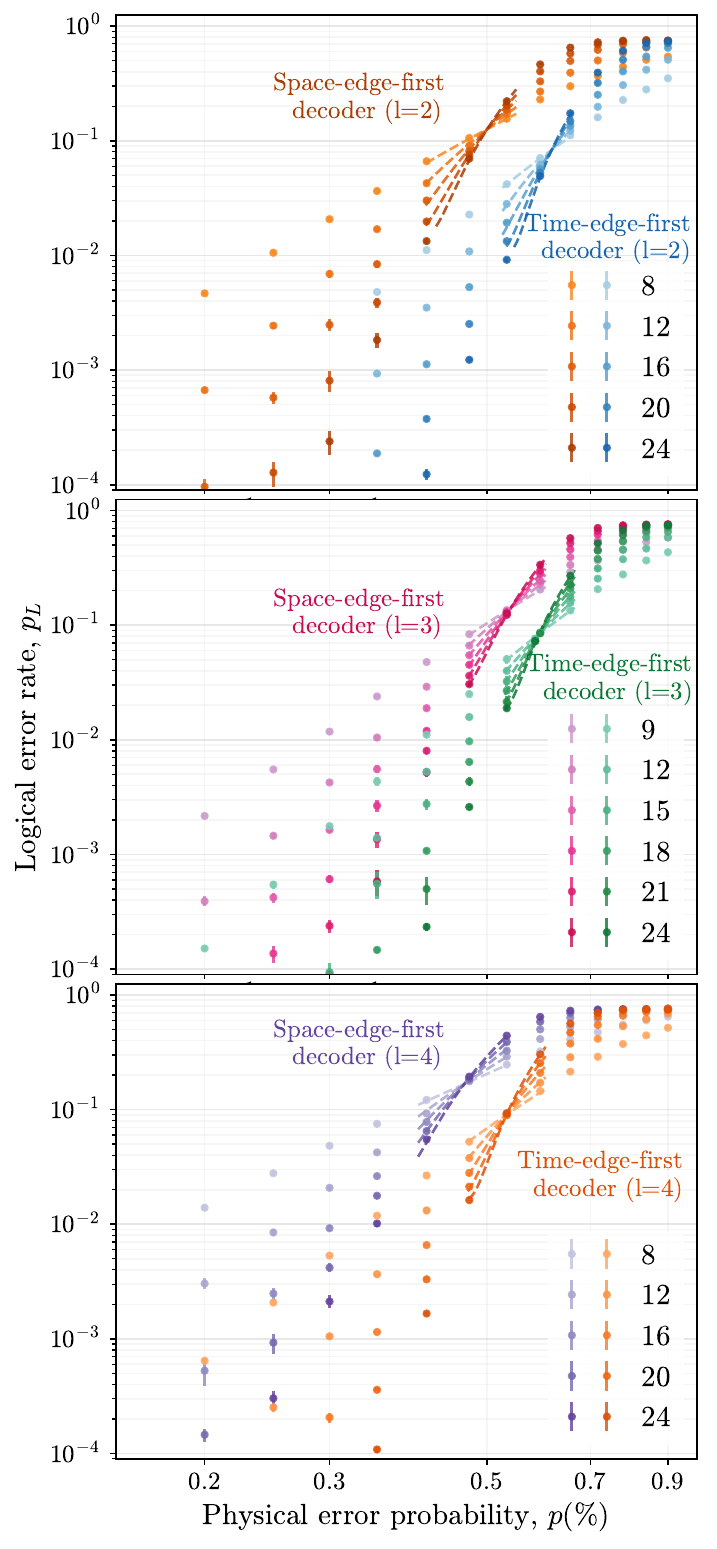}
    \caption{Comparison of circuit-level decoding performance under the time-edge-first and space-edge-first graph constructions for the fixed-width checks at different values of $l$. The legend indicates the code distance.}
    \label{fig:red_vs_comp_decoder}
\end{figure}

We present the numerical results for circuit-level simulation of $R=d$ rounds of syndrome extraction using decoders constructed from time-edge-first decomposition and the space-edge-first decomposition in Fig.~\ref{fig:red_vs_comp_decoder}.
Because $O(d)$ rounds of syndrome extraction are included in decoding, the effect of time distance is negligible.
For all $l$ we simulate, decoding using time-edge-first decomposition typically gives a higher threshold and lower logical rate. 
The threshold values for all cases are summarized in Table~\ref{tab:thresho}.
This indicates that the loss of the correlation information between detector flips is more significant in the space-edge-first decomposition.
We expect some improvement in the decoding method can further boost the decoding performance.

\begin{table}[htbp!]
    \centering
    \begin{tabular}{c|c|c|c}
        \toprule
        \diagbox{Decoder type}{Patch size} & $l=2$ & $l=3$ & $l=4$  \\
        \midrule
        Space-edge first & $0.50\%$ & $0.54\%$ & $0.46\%$ \\
        \midrule
        Time-edge first & $0.61\%$ & $0.59\%$ & $0.53\%$\\
        \bottomrule
    \end{tabular}
    \caption{Threshold values obtained using space-edge-first and time-edge-first decoders for patch sizes \(l = 2, 3, 4\).}
    \label{tab:thresho}
\end{table}

For the time-edge-first decomposition, the threshold decreases slightly as $l$ increases.
For $l=2$ and $l=3$, the decoding performance is quite similar.
However, the logical error rate increases dramatically when $l$ increases to $l=4$. 
This increase is due to higher average weights of checks in $l=4$ that leads to various internal errors and the correlation information loss during the hyperedge decomposition. 

In contrast, when the decoding graph is constructed from space-edge-first decomposition, the threshold increases and the logical error rate decreases slightly from $l=2$ to $l=3$.
This can be attributed to the fact that while both measurement schemes share the same time distance, checks with $l=3$ retains more single-shot structure and has sparser time edges and space-time edges. 
However, the performance again decreases at $l=4$ due to the higher frequency of high-degree hyperedges requiring auxiliary detectors. 
See App.~\ref{App:decomp} for a detailed discussion on hyperedge decomposition.

\subsection{Sliding-window decoding using different window sizes and patch sizes}\label{App:circuit_l_compare}

\begin{figure*}[htbp!]
    \centering
    \includegraphics[width=0.99\linewidth]{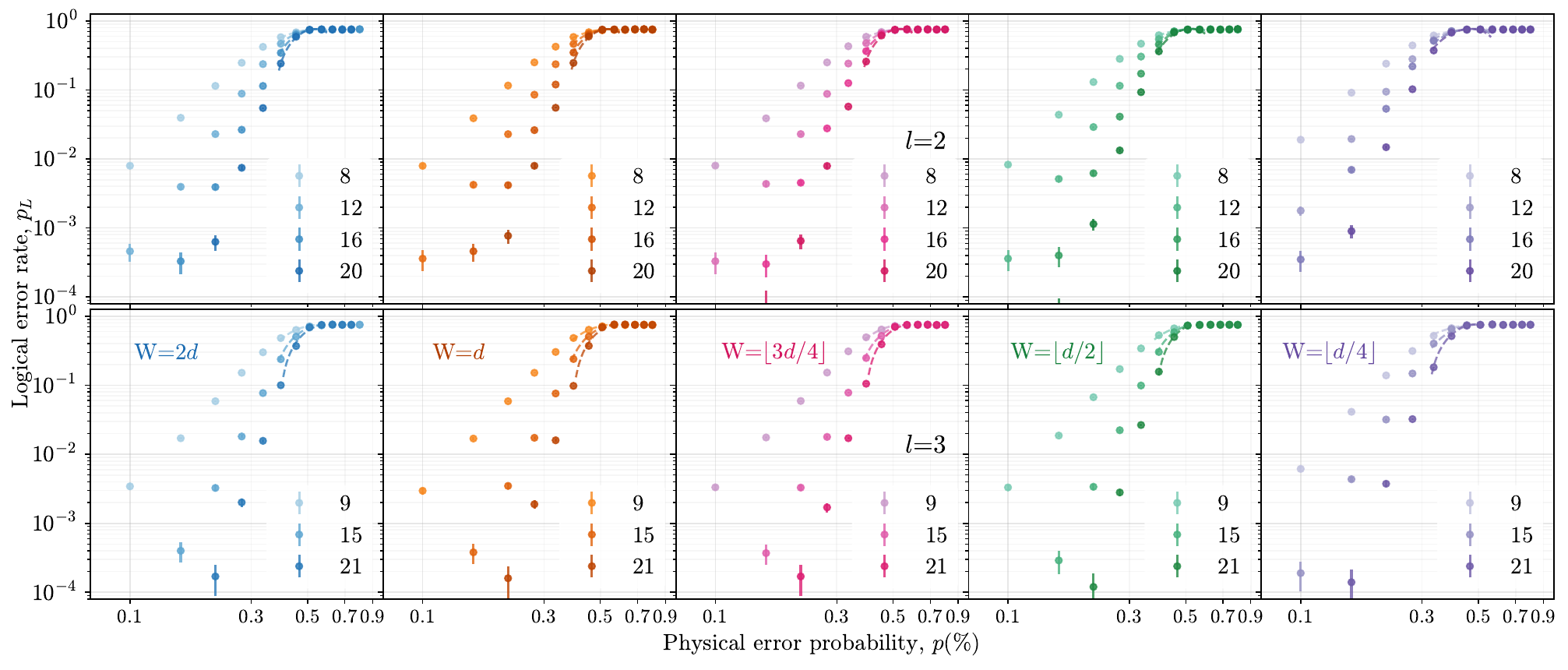}
    \caption{Comparison of logical error rates under circuit-level noise for sliding-window decoding with different window sizes $W$ and patch sizes $l=2,3$. The legend indicates the code distance.}
    \label{fig:Patch_size_comparision}
\end{figure*}

We present the performance of local single-shot checks of different patch sizes using the offset measurement scheme with sliding window of different window sizes $W$ under the circuit-level simulation with a decoder constructed with the space-edge-first decomposition in Fig.~\ref{fig:Patch_size_comparision}. 
For both $l=2$ and $3$, the decoding performance is similar for $W$ above $\lfloor d/2\rfloor$ in the region we simulate. 
However, the logical error rate increases when $W=\lfloor d/4\rfloor$.

Additionally, we observe that increasing the patch size from $l=2$ to $l=3$ results in a higher threshold value, similar to the behavior seen in Fig.~\ref{fig:circuit_results}, where we use $R=d$ rounds of syndrome extraction.
The corresponding threshold values are listed in Table~\ref{tab:threshold}.

\begin{table}[htbp!]
    \centering
    \begin{tabular}{c|c|c|c|c|c}
        \toprule
        \diagbox{Patch Size}{Window Size} & W=$2d$ & W=$d$ & $W=\lfloor 3d/4 \rfloor$ & W=$\lfloor d/2 \rfloor$ & W=$\lfloor d/4 \rfloor$ \\
        \midrule
        $l=2$ & $0.51\%$ & $0.52\%$ & $0.51\%$ & $0.50\%$ & $0.41\%$\\
        \midrule
        $l=3$ & $0.56\%$ & $0.58\%$ & $0.56\%$ & $0.56\%$ & $0.46\%$\\
        \bottomrule
    \end{tabular}
    \caption{Threshold values for the fixed-width local single-shot check for different patch sizes, evaluated using a sliding-window decoder with various window sizes.}
    \label{tab:threshold}
\end{table}

\end{document}